\documentclass[journal, twoside]{IEEEtran}
\usepackage{cite}
\usepackage{url}
\usepackage{amsmath,amssymb,amsfonts}
\usepackage{algorithmic}
\usepackage{graphicx}
\usepackage{dblfloatfix}
\usepackage{textcomp}
\usepackage{siunitx}
\usepackage{array}
\usepackage{float}
\usepackage{color}
\usepackage{placeins}
\usepackage{tabularx,colortbl}
\usepackage{balance}
\usepackage{subfigure}
\usepackage{booktabs}
\usepackage{multirow}
\usepackage{xcolor}
\usepackage{bm}
\usepackage{indentfirst}
\usepackage{multicol}
\usepackage{mathrsfs}
\usepackage{hyperref}
\graphicspath{{./Figs/}}

\usepackage{dsfont}

\usepackage{threeparttable}

\newcommand{\Rmnum}[1]{\expandafter\@slowromancap\romannumeral #1@}
\makeatother
\normalsize

\begin{document}
\title{On the Feasibility of Out-of-Band Spatial Channel Information for Millimeter-Wave Beam Search}
\author{
	Peize Zhang, \IEEEmembership{Member, IEEE},
	Pekka Ky\"{o}sti,
	Katsuyuki Haneda, \IEEEmembership{Member, IEEE},\\
	Pasi Koivum\"{a}ki, \IEEEmembership{Member, IEEE},
	and Wei Fan, \IEEEmembership{Senior Member, IEEE}
	\thanks{Manuscript received \qquad\qquad; revised \qquad\qquad; accepted \qquad\qquad. Date of publication  \qquad\qquad; date of current \qquad. This work was partly supported by the European Commission through the H2020 project Hexa-X (Grant Agreement no.\ 101015956) and partly by 6G Flagship programme, funded by Academy of Finland (Grant no.\ 318927). \emph{(Corresponding author: \qquad.)}}
	\thanks{Peize Zhang is with the Centre for Wireless Communications, University of Oulu, 90570 Oulu, Finland (e-mail: peize.zhang@oulu.fi).}
	\thanks{Pekka Ky\"{o}sti is with the Centre for Wireless Communications, University of Oulu, 90570 Oulu, Finland, and also with Keysight Technologies Finland Oy, 90630 Oulu, Finland (e-mail: pekka.kyosti@oulu.fi).}
	\thanks{Katsuyuki Haneda and Pasi Koivum\"{a}ki are with the Department of Electronics and Nanoengineering, Aalto University, 02150 Espoo, Finland (e-mail: katsuyuki.haneda@aalto.fi; pasi.koivumaki@aalto.fi).}
	\thanks{Wei Fan is with the Antenna, Propagation and Millimeter-Wave Systems Section, Department of Electronic Systems, Faculty of Engineering and Science, Aalborg University, 9220 Aalborg, Denmark (e-mail: wfa@es.aau.dk).}
	\thanks{Color versions of one or more of the figures in this paper are available online at http://ieeexplore.ieee.org.}
	\thanks{Digital Object Identifier \qquad\qquad}
}

\markboth{IEEE Transactions on Antennas and Propagation}{Ky\"{o}sti \MakeLowercase{\textit{et al.}}: On the Feasibility of Out-of-Band Spatial Channel Information for Millimeter-Wave Beam Search}

\maketitle
\boldmath
\begin{abstract}
The rollout of millimeter-wave (mmWave) cellular network enables us to realize the full potential of 5G/6G with vastly improved throughput and ultra-low latency. MmWave communication relies on highly directional transmission, which significantly increase the training overhead for fine beam alignment. The concept of using out-of-band spatial information to aid mmWave beam search is developed when multi-band systems operating in parallel. The feasibility of leveraging low-band channel information for coarse estimation of high-band beam directions strongly depends on the spatial congruence between two frequency bands. In this paper, we try to provide insights into the answers of two important questions. First, how similar is the power angular spectra (PAS) of radio channels between two well-separated frequency bands? Then, what is the impact of practical system configurations on spatial channel similarity? Specifically, the beam direction-based metric is proposed to measure the power loss and number of false directions if out-of-band spatial information is used instead of in-band information. This metric is more practical and useful than comparing normalized PAS directly. Point cloud ray-tracing and measurement results across multiple frequency bands and environments show that the degree of spatial similarity of beamformed channels is related to antenna beamwidth, frequency gap, and radio link conditions.
\end{abstract}
\unboldmath


\section{Introduction}
\IEEEPARstart{A}{s the} global rollout and adoption of 5G technology continue, next-generation wireless communication systems (i.e., B5G/6G) are expected to provide ultra-high throughput and seamless connectivity in coexisting sub-6 GHz and millimeter-wave (mmWave) cellular networks~\cite{Tataria20216G}. From the propagation characteristics analysis perspective, high-capacity mmWave communications will likely offer up to a few hundred meters of directed coverage~\cite{Zhang2020Meas}, which becomes much shorter in non-line-of-sight (NLOS) scenario~\cite{Zhang2021Radio}, because of severe path loss and sensitivity to link blockages. Apart from the frequency-dependent free-space path loss, the efficiency of diffraction and capacity of penetration also strongly decrease with increasing frequency~\cite{Shafi18Microwave}. Fortunately, sub-6 GHz radio systems can establish more flexible and reliable communication links with continuous coverage, which benefits from relatively shorter wavelength.

It is known that the antenna gain has to be significantly increased to compensate the severe propagation loss at higher frequency, while higher antenna gain means practically more directivity and narrower beamwidth of antenna radiation pattern. Moreover, initialization of a communication link necessitates a beam search procedure from tranceivers relying on adaptive high-gain antenna systems~\cite{Rag2016Beam}. The high-gain antenna, which transmits a much narrower beam, is therefore susceptible to loss of signal, leading to a more laborious beam searching scheme~\cite{Giordani2019Tut}. Modern radio transceivers are capable of supporting multi-band cooperation transmission with tens or even hundreds of antenna elements. Therefore, it is attractive to reuse out-of-band spatial channel information (e.g., low- and mid-bands) for fast higher frequency (e.g., mmWave band) beam steering and alignment with reduced training overhead~\cite{Nuria2017Millmeter,Ali2018Millimeter}.

To this end, the primary task is explore how similar the propagation channels are at separate frequency bands and how much useful spatial channel information one band can provide to the other. With regards to the standardization of wave propagation models, frequency dependence of channel parameters has been a topic of interest for many years\cite{3GPP_38.901}. Channel models can not only be significantly simplified if the radio signals can propagate in the similar manner across difference bands, but also be used to predict channel characteristics if new frequency bands are considered for radio communications. Furthermore, channel similarity across different frequency bands can be exploited to improve system performance. For example, the method proposed in \cite{refA} performs rough and quick angle estimation at a lower frequency and reuses the estimated spatial profiles for precious beam operation at high frequency, considering the high similarity of spatial profiles between low and high frequency bands. Extensive empirical analyses of indoor and outdoor spatial channels have been promoted, where measured power-angle-delay profiles (PADPs) in the identical transmitter-receiver (T-R) combinations are presented and compared over multiple frequency bands from microwave to mmWave bands\cite{refA,refB,Zhang2018Indoor,Zhang2019Millimeter,refF}. Analysis of directional propagation channels reveals that similar dominant clusters (i.e., paths) can be observed in different frequency bands, yet many more weak multipath components (MPCs) can be detected in relatively lower bands compared to the results in higher bands. Meanwhile, it is clear that spatial profiles at mmWave bands tend to be more specular and sparse. In addition, path loss and delay dispersion parameters at 28, 73, and 140 GHz in an indoor scenario have been presented, where good similarity in path loss yet reduced time dispersion were observed in higher frequency bands\cite{refC}. In \cite{refE}, channel sounding results reported in mmMAGIC project (covering different propagation scenarios and different frequency bands in 2 to 86 GHz\cite{mmMAGICD22}) were summarized with a focus on root mean square (rms) delay spread (DS), showing that any frequency trend of the DS is small considering its confidence intervals and largely dependent on the specific scenario.

Existing studies mainly focus on visual comparison of spatial-temporal channel characterization for the specific T-R combination at different frequencies, along with the statistical analysis of channel parameters based on a large amount of propagation data collected across different environments. Though multi-band channel similarity can be observed, there is a dearth of standard metrics to characterize the degree of similarity among low-, mid-, and high-bands. To the best of our knowledge, except for the metric derived from the multipath fading trends with respect to the power delay profile (PDP) as reported in \cite{Yi2021Multipath}, no definite metric has been proposed to measure the spatial channel similarity in terms of power angular spectra (PAS). With the number of antenna elements and inter-element spacing increasing, the angular resolution of radio channel will increase accordingly. It is also essential to consider practical antenna size limitations when leveraging low-band spatial channel information for high-band beam search. Thus, the aims of this paper are twofold:
\begin{itemize}
	\item The first is to develop a metric for the evaluation of spatial channel similarity at two different frequencies under the practical antenna constrains (i.e., antenna radiation pattern). To conduct the statistical analysis, we need to collect a sufficient large amount of multi-band propagation channel data at the same transmitter (Tx) and receiver (Rx) locations. Use of point clouds for ray-tracing channel simulation in site-specific environment is considered, which is an attractive approach given the difficulty of conducting extensive channel measurement campaigns consisting of thousands of T-R links. Meanwhile, dual-band channel sounding results in different environments are also used for validation. Consequently, two methods are proposed to analyze the spatial channel similarity, where the former offers a probability measure on the total variation distance of PAS while the latter offers a more practical and intuitive scheme to identify the available beam directions.
	\item The second is to provide valuable insights into the feasibility of out-of-band (e.g., low- and mid-bands) spatial channel information for mmWave beam steering and alignment. Based on the beam direction-based channel similarity measure, sets of best beam directions at two frequencies can be obtained and utilized for calculating the number of useless directions with respect to the estimated result in lower band. Depending by the system configurations (including antenna configurations, channel conditions, and beam direction estimation methods), the feasibility of leveraging lower band channel information for mmWave beam search will change accordingly.
\end{itemize}

The remainder of this paper is organized as follows. Section \ref{Sec:RTData} details the point cloud ray-tracing simulations and channel measurements across multiple frequency bands and environments. Section \ref{Sec:Methods} describes two proposed metrics to measure the spatial channel similarity based on the multi-band PAS. Section \ref{Sec:RandD} discusses the feasibility of lower frequency spatial channel information for mmWave beam seach using practical antenna configurations. Finally, Section \ref{Sec:Conclusion} concludes the paper.

\section{Multi-Band Propagation Channel Data Collection}
\label{Sec:RTData}
\subsection{Point Cloud Ray-Tracing Simulations}
\begin{figure}[!t]
	\centering
	\subfigure[][]{
		\label{fig:pointcloud}
		\includegraphics[width=0.35\textwidth]{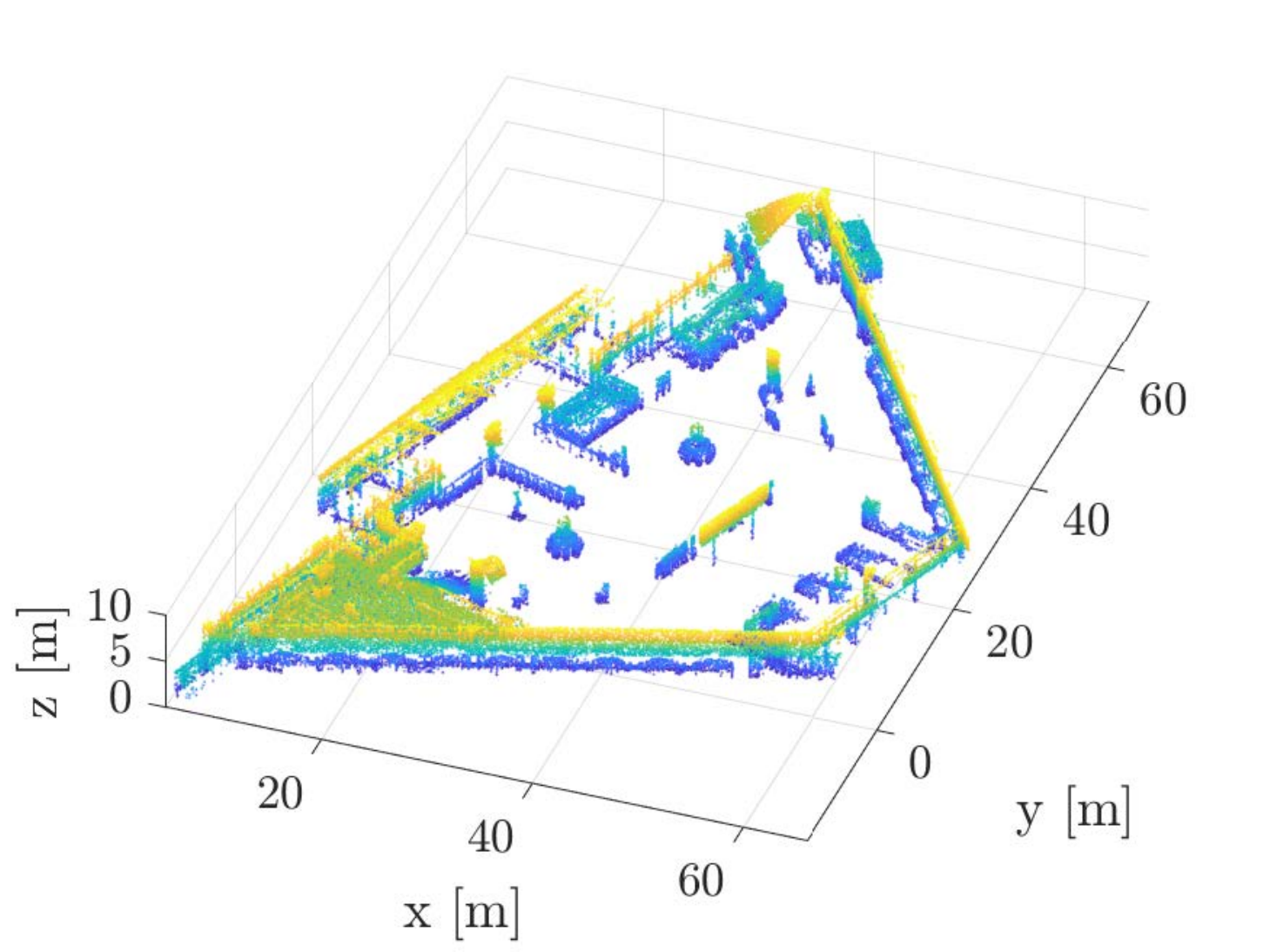}
	}
	\subfigure[][]{
		\label{fig:linkLoc}
		\includegraphics[width=0.25\textwidth]{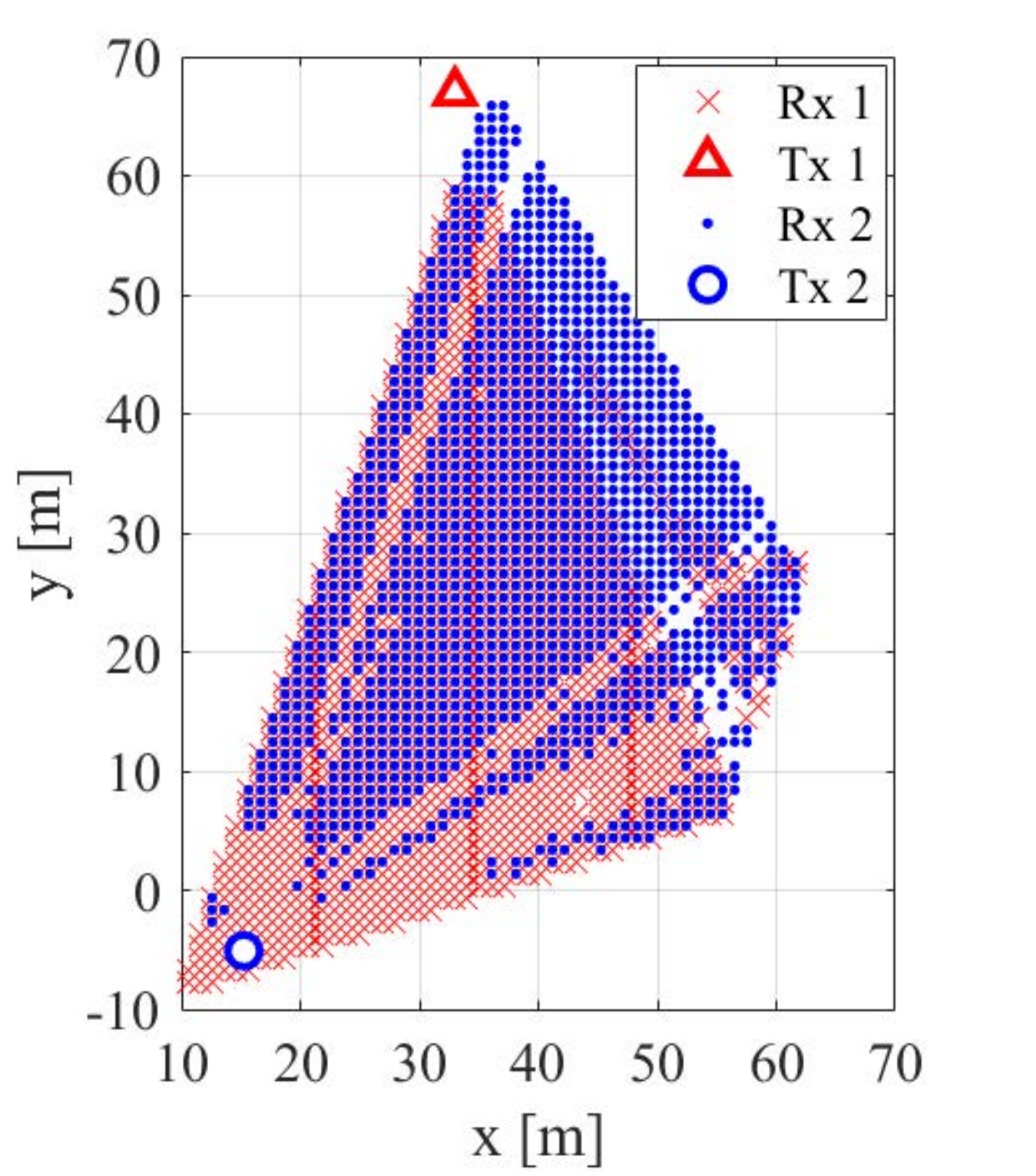}
	}
	\caption{(a) Laser-scanned point cloud used in ray-tracing, where the points are colored according to their position on the $z$-axis. (b) Tx and Rx positions.}
	\label{Fig:SimEnvi}
	\vspace{-6pt}
\end{figure}
The propagation channel data used in this work is generated via ray-tracing simulations in applying a laser-scanned point cloud model of the Helsinki-Vantaa airport check-in area at Terminal~$2$\cite{Koivumaki_TAP21,Jarvelainen16_TAP}. The raw point cloud is obtained with a Z+F IMAGER$^\circledR$ 5006h 3D laser-scanner\cite{scanner}, which uses movable mirrors to steer a laser beam in different directions to detect the distances to reflective surfaces.  The point cloud used in ray-tracing simulations is illustrated in Fig.~\ref{fig:pointcloud}. Note that the raw point cloud is preprocessed by removing the ceiling of the airport terminal (i.e., points above $10$~m), as well as the floor, which will significantly reduce the total number of points for ray-tracing simulations. Other clutters, for example, pedestrians that were present during the laser scanning, are removed so that the ray-tracing simulations were performed without any human present in the hall. The points that represent the walls of the airport terminal, check-in kiosks, and other fixtures of the environment are kept during ray-tracing simulations. As shown in Fig.~\ref{fig:linkLoc}, two Tx locations are selected, where Tx1 is in a side corridor and Tx2 is behind the stairs in the check-in area. Two Rx sets are selected, including thousands of Rx locations, to establish communication links to Tx1 and Tx2, respectively. Propagation data contains in total 2875 links, composed of 1473 links between Tx1 and Rx1 and 1402 links between Tx2 and Rx2. It can be observed that Rx locations are not fully uniform since certain sectors with direct paths to the corresponding Rx locations are excluded. Thus, only NLOS links are considered during the simulation, while line-of-sight (LOS) links are all blocked by physical objects.

As regards to the ray-tracing simulation setup, the following propagation mechanisms are considered in the simulations, including single- and double-bounce paths, diffracted paths, and diffraction to single and double-bounce paths. Total five frequency bands (i.e., 4, 15, 28, 60, and 86 GHz) are considered. The permittivity of the environment and shadowing losses for each frequency, which were used in the ray-tracing simulations, are summarized in Table~\ref{tab:rt_params}. These parameters are optimized based on the channel measurement campaign conducted at the airport~\cite{Vehmas16_VTC} folowing the methodology outlined in~\cite{mmMAGICD22}. Note that the measured or optimized parameters were not available at $4$~GHz, and they were obtained using a linear fit according to the available parameters in other bands. The permittivity of the physical objects in the environment is assumed to be identical for simplicity of the measurement-based calibration and is used to calculate reflection and diffraction coefficients. A flat shadowing loss calibrated based on field measurement data is applied to the rays that are \textit{shadowed}, i.e., its corresponding $1^{\rm st}$ Fresnel ellipsoid is not free of obstructions~\cite{Jarvelainen_letter16,Koivumaki_TAP22}.
\begin{table}[!t]
	\centering
	\caption{Ray-tracing simulation parameters.}
	\label{tab:rt_params}
	\begin{tabular}{l|c c c c c}
		\hline\hline
		Frequency [GHz]& $4$ & $15$ & $28$ & $60$ & $86$\\\hline
		Permittivity &  $4.3$ & $4.2$ & $3.9$ & $3.6$ & $3.3$\\\hline
		Shadowing loss [dB]& $20$ & $30$ & $60$ & $100$ & $130$\\
		\hline\hline
	\end{tabular}
\vspace{-6pt}
\end{table}

\subsection{Channel Measurement Campaigns}
\begin{figure}[!t]
	\centering
	\subfigure[][]{
		\label{fig:Corridor}
		\includegraphics[height=1.4in]{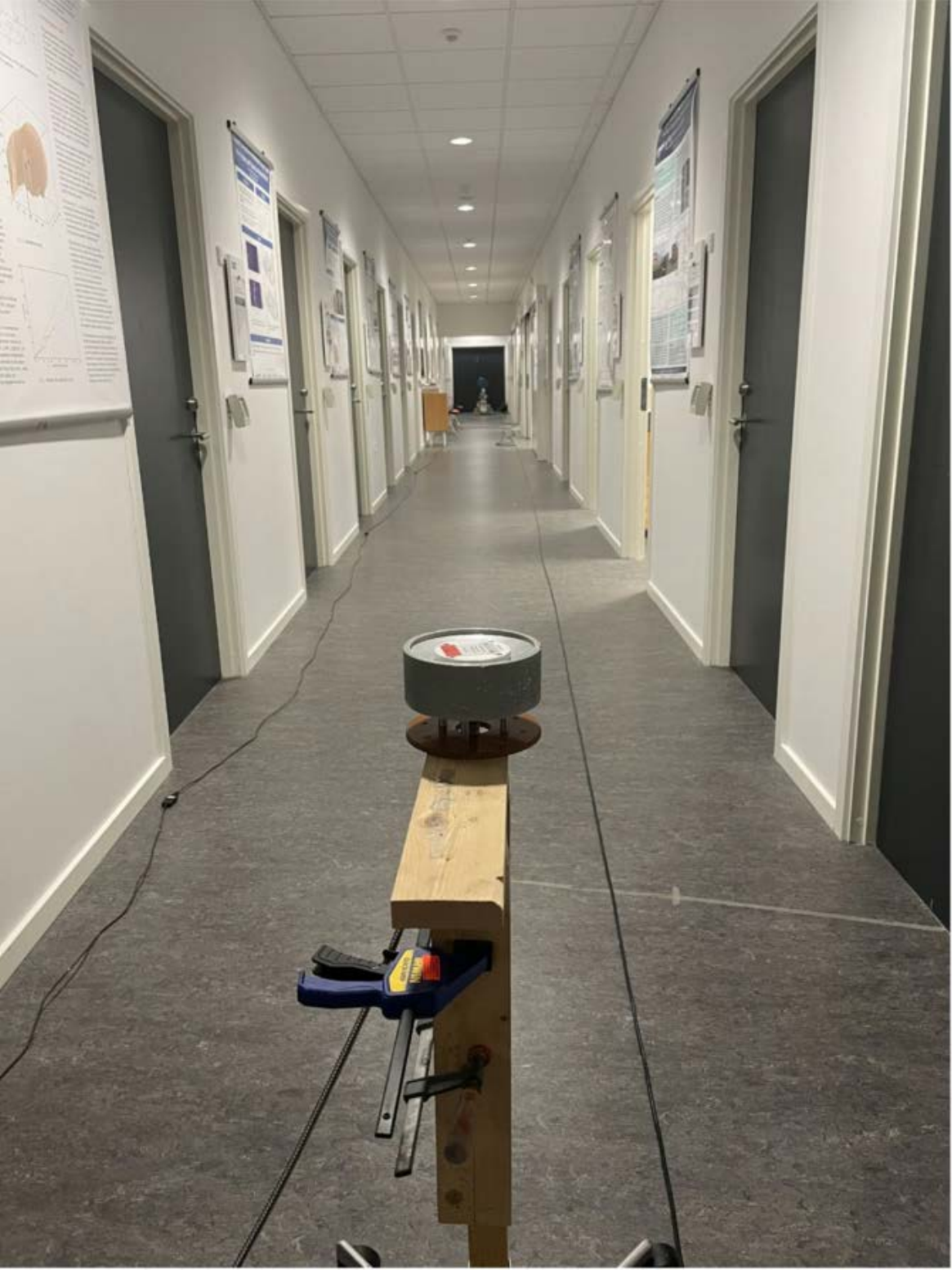}
	}
	\subfigure[][]{
		\label{fig:Entrance}
		\includegraphics[height=1.4in]{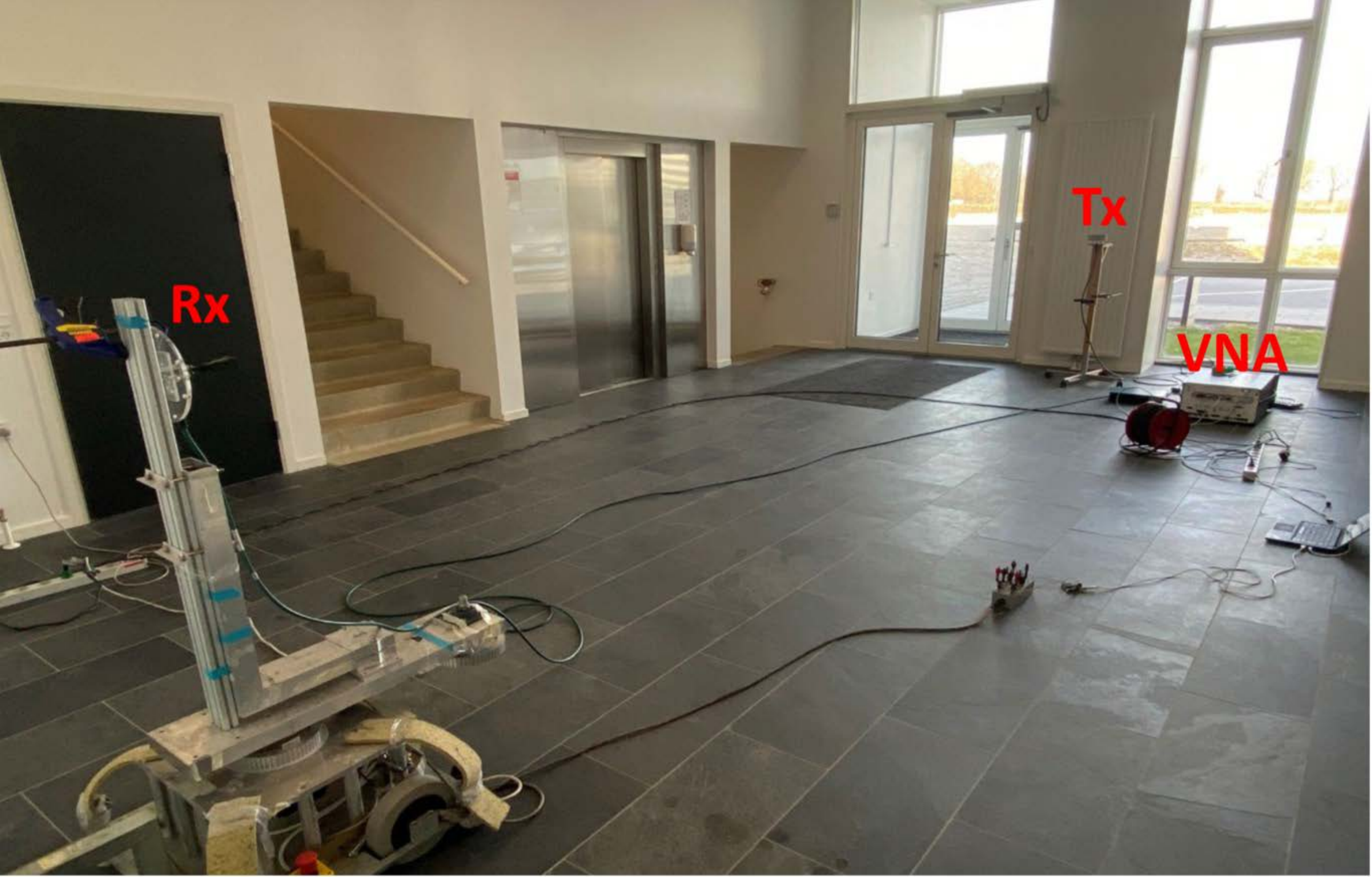}
	}
	\caption{Channel measurement campaigns in (a) indoor corridor and (b) entrance hall.}
	\label{Fig:MeasEnvi}
	\vspace{0pt}
\end{figure}
Channel measurements in two indoor scenarios (see Fig. \ref{Fig:MeasEnvi}) were performed in Aalborg University, Denmark. A long-range vector network analyzer (VNA) based channel sounding system enabled by radio over fiber (RoF) concept was employed for the measurements. The details about the channel sounder at 28 GHz can be found in \cite{Mbugua2022Phase}. In \cite{Lyu2021Design}, a long-range channel sounder at 220--330 GHz was reported. Same concept is employed for the 100 GHz channel sounder utilized in our measurement, except that different frequency extenders are used.  For both channel measurements, omnidirectional antennas were used at the Tx side, while directional antennas were employed at the Rx side to cover 360$^\circ$ in the azimuth domain. Directional channel sounding scheme (DSS) was employed at the Rx side to measure the spatial channel profile directly with the help of the turntable. Note that to ensure that channel spatial profiles at different frequency bands are not affected by the antenna pattern, directional antennas with similar half-power beamwidth (HPBW) were employed. The measured PADPs are estimated based on the method proposed in \cite{Li2022Virtual}, which can be further used to extract MPCs with peak detection algorithm. A dynamic range of 30 dB is set in the measurement results.
\begin{table}[!t]
	\centering
	\caption{Antenna configurations for dual-band channel measurements.}
	\label{tab:AntConfigure}
	\begin{tabular}{l|c c|c c}
		\hline\hline
		Envi. &  \multicolumn{2}{c|}{Indoor corridor} & \multicolumn{2}{c}{Entrance hall}\\
		\hline
		Freq. range [GHz] & 28--30 & 99-101 & 28--30 & 99-101 \\
		\hline
		Freq. point & \multicolumn{2}{c|}{1001} & 1001 & 601 \\
		\hline
		Tx ant. type & \multicolumn{4}{c}{Omni.}\\
		\hline
		Rx ant. type & \multicolumn{4}{c}{Horn}\\
		\hline
		RX ant. height [m] & 0.93 & 1.25 & \multicolumn{2}{c}{1.25} \\
		\hline		
		RX ant. HPBW [deg] & 40 & 40 & 20.8 & 16 \\
		\hline		
		RX ant. Gain [dBi] & 13.5 & 13.5 & 19.5 & 20.5 \\
		\hline
		Rx rot. Step [deg] & 1.5 & 1 & 8 & 8 \\
		\hline\hline
	\end{tabular}
\vspace{0pt}
\end{table}

\begin{figure}[!t]
	\centering
	\subfigure[][]{
		\label{fig:PADP28}
		\includegraphics[width=3in]{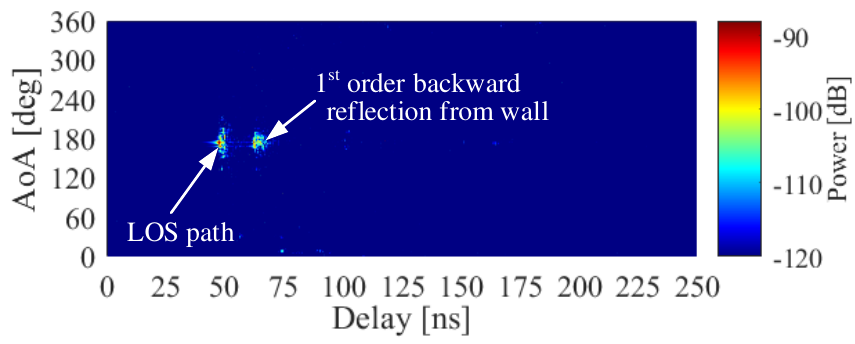}
	}
	\subfigure[][]{
		\label{fig:PADP100}
		\includegraphics[width=3in]{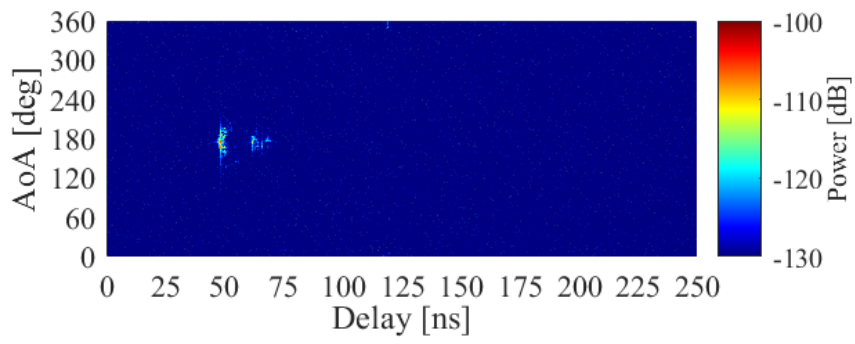}
	}
	\caption{Measured PADPs in indoor corridor at (a) 28 GHz and (b) 100 GHz.}
	\label{Fig:MeasPADP}
	\vspace{0pt}
\end{figure}

Only one LOS link in each environment was selected due to highly time-consuming measurements. For the indoor corridor scenario, a measurement bandwidth of 2 GHz with 1001 frequency points was set in the VNA, resulting in a delay resolution of 0.5 ns and maximum excess delay of 500 ns. The measurement distance was set to 14 m. The measured PADPs at 28 GHz and 100 GHz are shown in Fig. \ref{Fig:MeasPADP}. The measured channels are highly sparse and specular, and the same dominant paths along LOS directions can be identified at two frequencies. For the entrance hall scenario, 1001 and 601 frequency points were selected with the same sounding bandwidth of 2 GHz at 28 GHz and 100 GHz, respectively. The measurement distance was set to 4.84 m. The details of antenna configurations are summarized in Table \ref{tab:AntConfigure}.

\subsection{Multipath Channel Characterization}
The outputs of the ray-tracing simulations and channel measurements include all possible rays (i.e., MPCs) within a specific power range, where each ray is characterized by its received power, propagation delay, angle of arrival (AoA), and angle of departure (AoD) (if possible). Hence, the dicrete PADP can be written as $P(q;\Omega,\tau)$ if only consider the power distribution in AoA domain, where $q$ denotes the index for T-R combination, $\Omega$ denotes AoA, and $\tau$ denotes propagation delay. In this paper we only analyze the spatial channel similarity on Rx side and the same applies to Tx side relying on the reciprocity of propagation channel. For the sake of simplicity, we use sub-script $L$ and $U$ to denote the lower and upper frequency antenna systems, propagation channels, and their parameters, respectively. Consequently, each PADP can be represented as a sum of discrete paths $p=1,\dots,l$ as
\begin{equation}\label{eq:PAPD}
\begin{split}
	\left\{
	\begin{array}{ll}
		P_L(\Omega,\tau) = \sum\limits_{p=1}^{l_L} P_{L,p} \, \delta\left(\Omega - \Omega_{L,p}\right) \, \delta\left(\tau - \tau_{L,p}\right)\\
		P_U(\Omega,\tau) = \sum\limits_{p=1}^{l_U} P_{U,p} \, \delta\left(\Omega - \Omega_{U,p}\right) \, \delta\left(\tau - \tau_{U,p}\right),
	\end{array} \right .
\end{split}
\end{equation}
where link index $q$ is left out for brevity, $\delta(\cdot)$ is the delta function, $l$ is the total number of rays, $P_p$, $\Omega_p$, and $\tau_p$ are the received power\footnote{We assume unity transmit power and use \emph{received power} and \emph{channel gain} interchangeably.}, AoA, and propagation delay of the $p$th rays, respectively.

\begin{figure}[!t]
	\centering
	\includegraphics[width=0.3\textwidth]{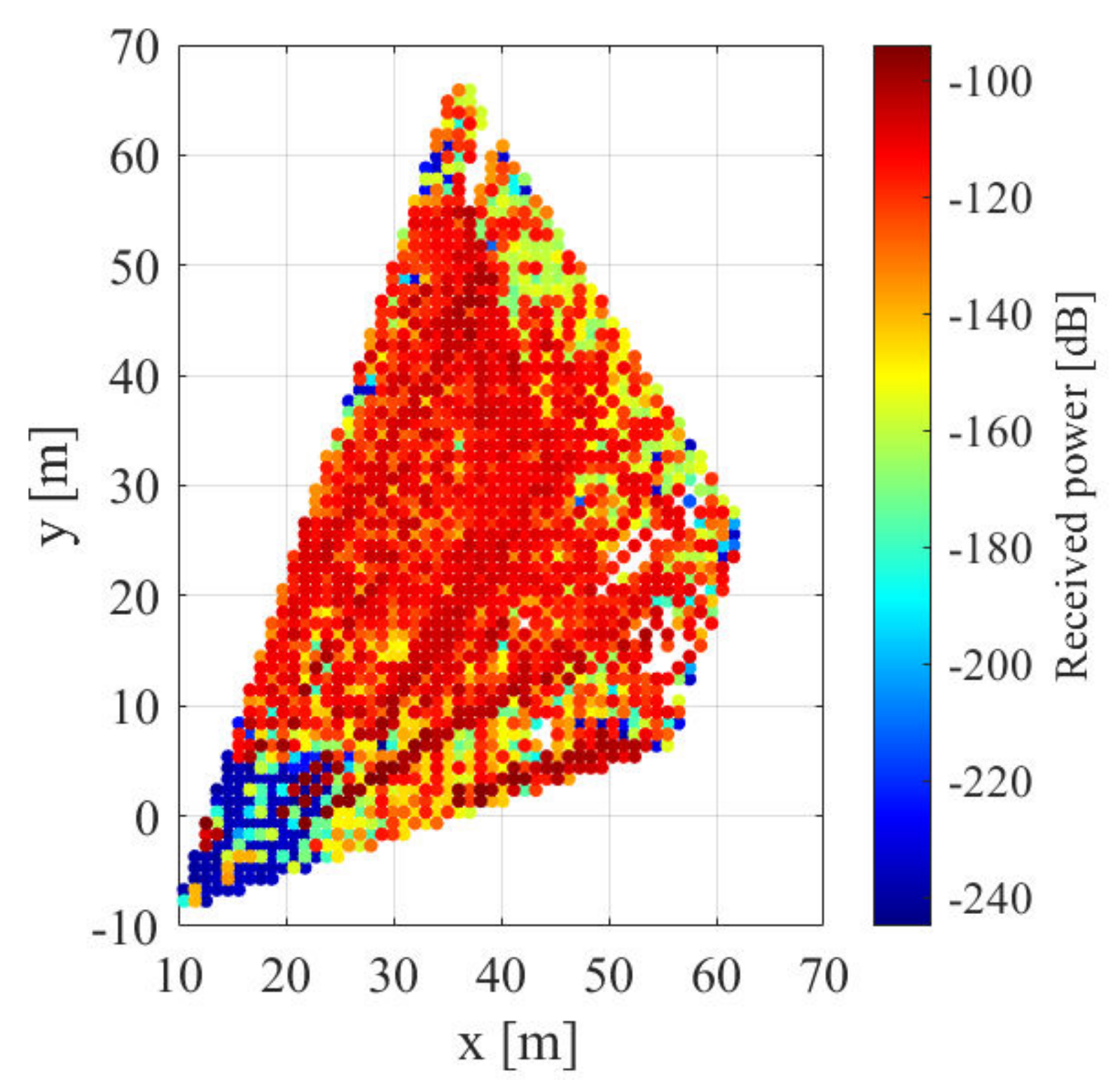}
	\caption{Total channel gain in each Rx position at 86 GHz.}\label{fig:linkGain}
	\vspace{0pt}
\end{figure}

\begin{figure}[!t]
\centering
\subfigure[][]{
	\label{fig:linkPAPDside}
	\includegraphics[width=0.21\textwidth]{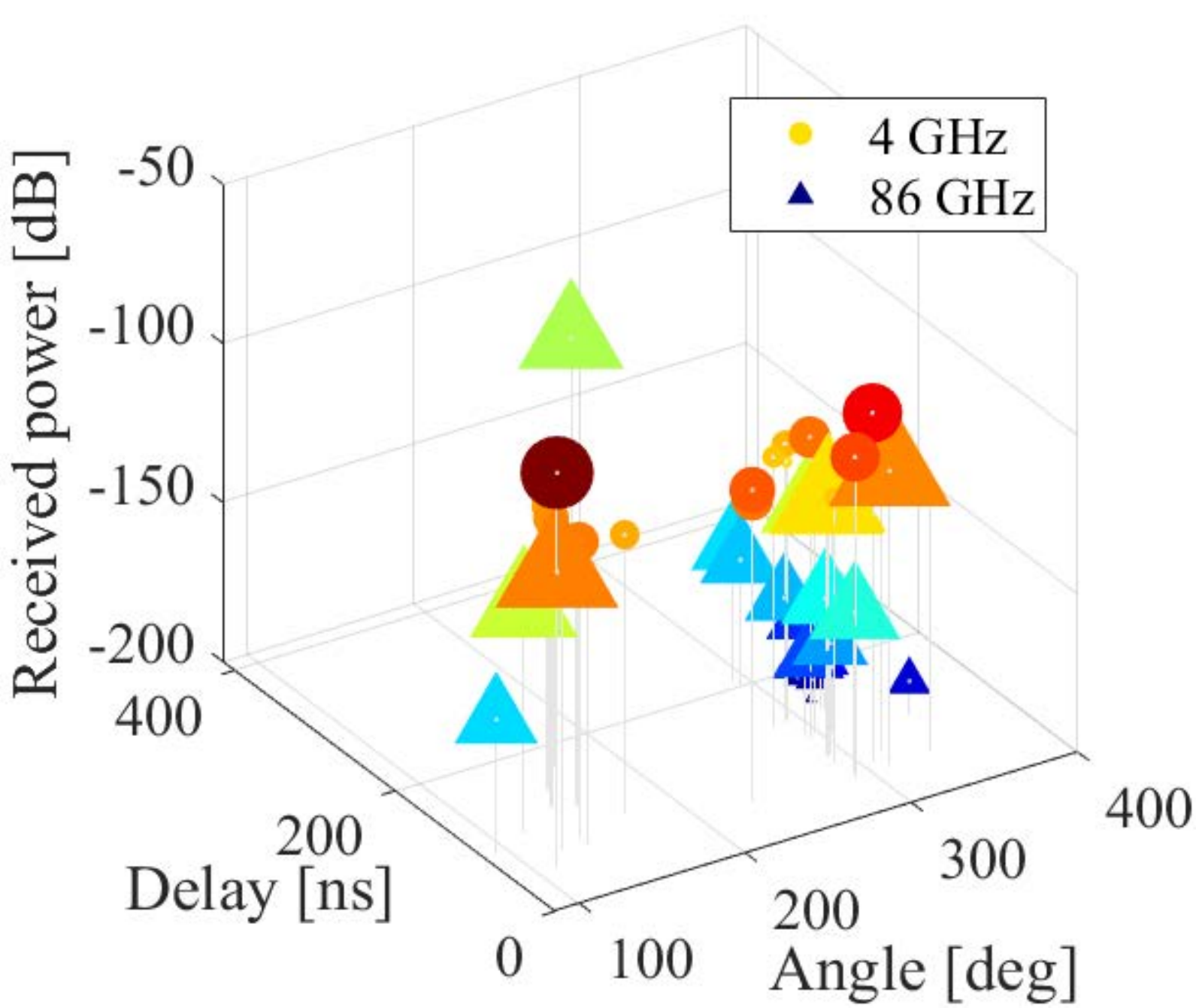}
}
\subfigure[][]{
	\label{fig:linkPAPDtop}
	\includegraphics[width=0.24\textwidth]{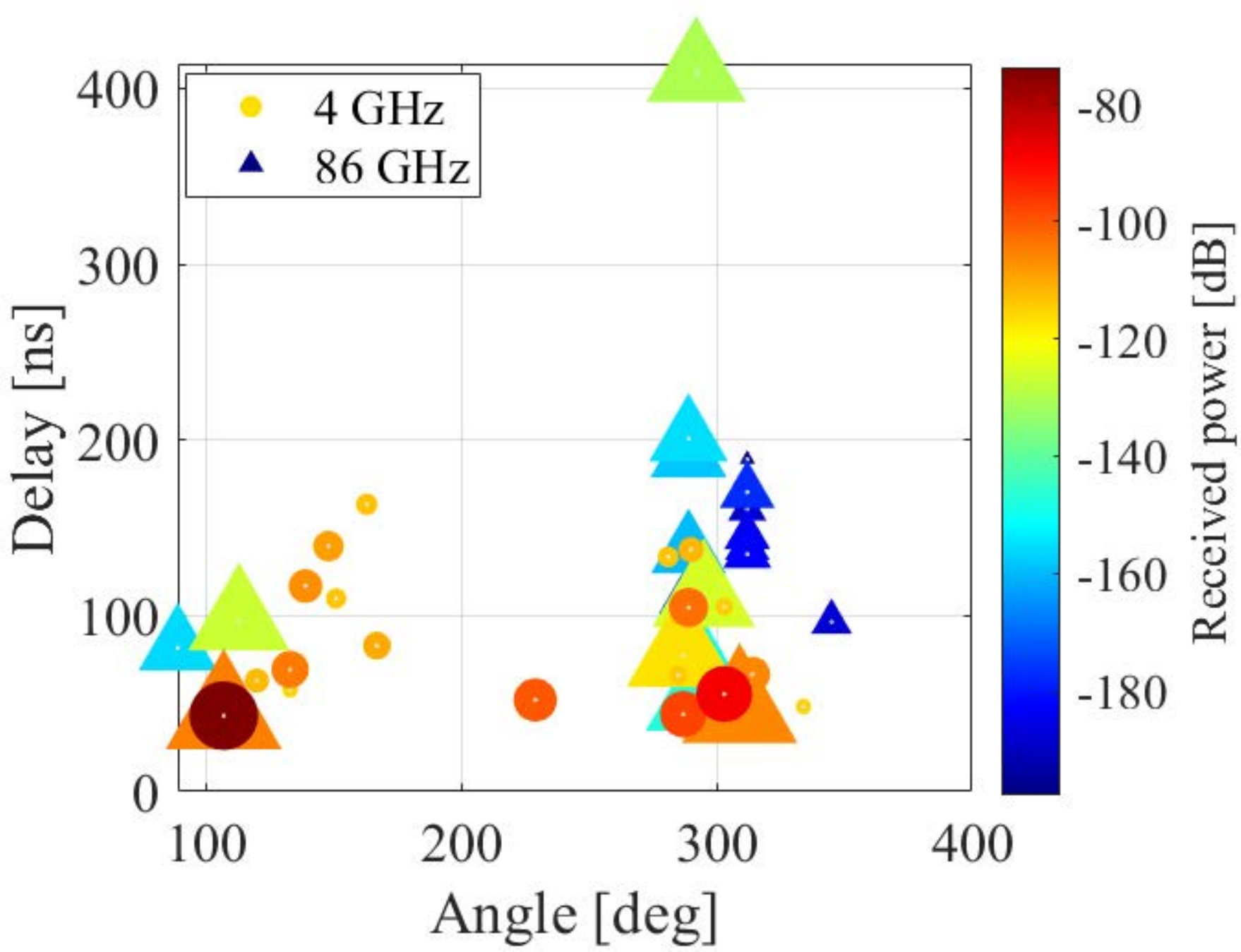}
}
\caption{Comparison of the PADPs at 4 GHz and 86 GHz for link 15. (a) 3D plot.  (b) Top view.}
\label{fig:linkPAPD}
\vspace{0pt}
\end{figure}

Fig. \ref{fig:linkGain} shows the total channel gain of each link at 86 GHz, which is calculated as the sum of all channel gains for every ray $\sum_{p=1}^{l}P_p$. A significantly large variation of close to 140 dB in channel gains among different Rx positions can be observed. Discrete PADPs of a specific link at 4 GHz and 86 GHz are shown in Fig. \ref{fig:linkPAPD}, where the circles and triangles represent the ray locations in low and high bands, respectively. In terms of the power difference, the $P_p$ at lower frequency is much larger than the results at higher frequency as indicated in Fig. \ref{fig:linkPAPDside}. It is a bit more difficult to explore the channel similarity visually without power normalization. On the other hand, from the top view of 3D PADP (see Fig. \ref{fig:linkPAPDtop}), dominant rays across two frequencies share similar propagation delays and AoAs. It means that the fading trend in delay and angular domains can be compared and defined as a metric for channel similarity, as well as the dominant propagation directions. Existing theoretical and empirical studies, in general, exploit the scheme that using low-band spatial channel information assists high-band beam steering and alignment. Henceforth, the 3D PADP will shrink to PAS for analysis of multi-band spatial channel similarity.

\section{Spatial Channel Similarity Metrics}
\label{Sec:Methods}
In this section, two metrics used for measuring the spatial channel similarity are proposed based on the multi-band propagation channel simulation data. The first method applies the widely known total variation distance of probability measures on the PAS estimated over different frequency bands. The second method is slightly more practical, which assumes certain antenna beam widths at different frequencies and tries to directly evaluate the usability of spatial channel information extracted at one frequency for beam search at the other frequency. Following the definitions of two metrics, the comparison study has been conducted across different frequency combinations and methods.

\subsection{PAS Similarity Percentage}
A intuitive view of multi-band spatial channel similarity is to compare the discrete and sparse PAS. However, it is difficult to promote direct comparison of PAS considering the received power difference between low and high frequencies. Thus, the idea is to first normalize the PAS and then develop a metric, so-called PAS similarity percentage (PSP), to measure the total variation distance of normalized PAS. According to procedures reported in \cite{Kyosti-18b}, the continuous PAS are first estimated by filtering the actual power angular distributions of the propagation channel by a function that corresponds to a limited aperture of an antenna. The filtering is performed for both of the considered frequencies. Consequently, the both spectra are normalized to the sum power of unity such that they can be interpreted as PDFs. Finally, the total variation distance between the probability distributions (i.e., normalized PAS) is calculated.
\begin{figure}[!t]
	\centering
	\includegraphics[width=2.8in]{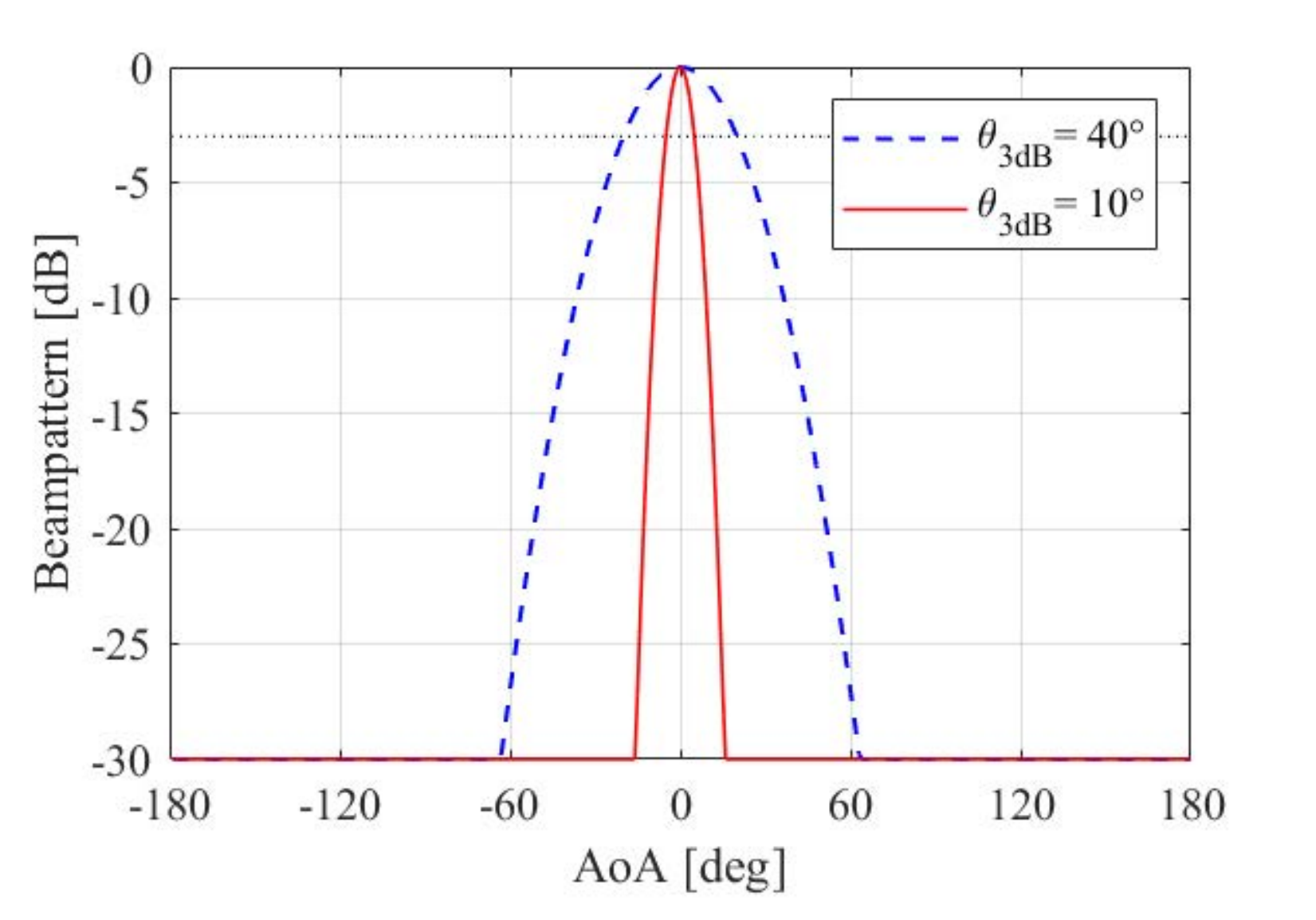}
	\caption{The normalized gain pattern of the 3GPP beam with $\theta_{\mathrm{3dB}}=$ 10$^{\circ}$ and 40$^{\circ}$.}\label{fig:3gppBeam}
	\vspace{-6pt}
\end{figure}

\subsubsection{Beam Filtering}
Any antennas or arrays can be used for defining the angular filter. It can be, for example, a phased array and the filtering operation would equal the classical Bartlett beamforming. Here, a simple synthetic radiation pattern $G(\Omega)$ specified by the 3GPP in \cite[Table 7.3-1]{3GPP_38.901} is adopted. As shown in Fig. \ref{fig:3gppBeam}, the peak to minimum gain ratio is 30 dB and 10$^{\circ}$ is chosen as the definable HPBW $\theta_{\mathrm{3dB}}$. We take convolution $\int P(\Omega) \, G(\alpha-\Omega) d\Omega$ of the propagation channel PAS $P(\Omega)$ and the angular filter $G(\Omega)$. This corresponds to steering the beam to a direction $\alpha$ and collecting the sum power from all observable MPCs weighted with corresponding beam gains. The resulting power with steering angle $\alpha$ is
\begin{equation}\label{eq:conv}
	\begin{split}
		\left\{ \begin{array}{ll}
			B_L(\alpha) =  \sum\limits_{p=1}^{l_L} P_{L,p} \, G(\alpha-\Omega_{L,p})\\
			B_U(\alpha) =  \sum\limits_{p=1}^{l_U} P_{U,p} \, G(\alpha-\Omega_{U,p}).
		\end{array} \right .
	\end{split}
\end{equation}
\begin{figure}[!t]
	\centering
	\subfigure[][]{
		\label{fig:PSP_2}
		\includegraphics[width=3.2in]{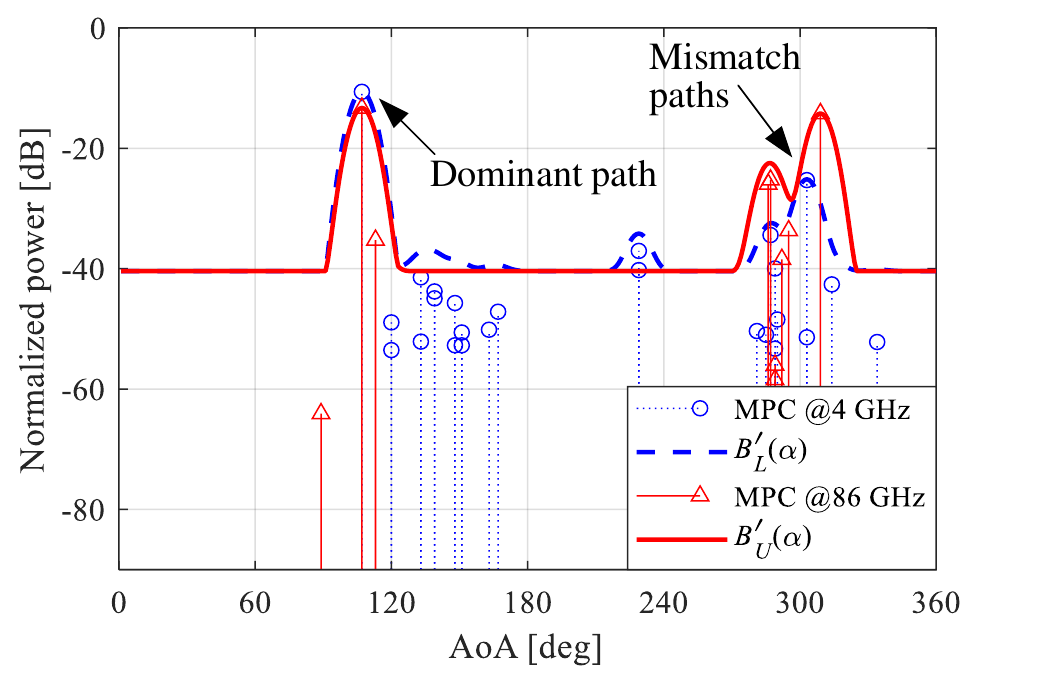}
	}
	\subfigure[][]{
		\label{fig:PSP_all_cdf}
		\includegraphics[width=3.2in]{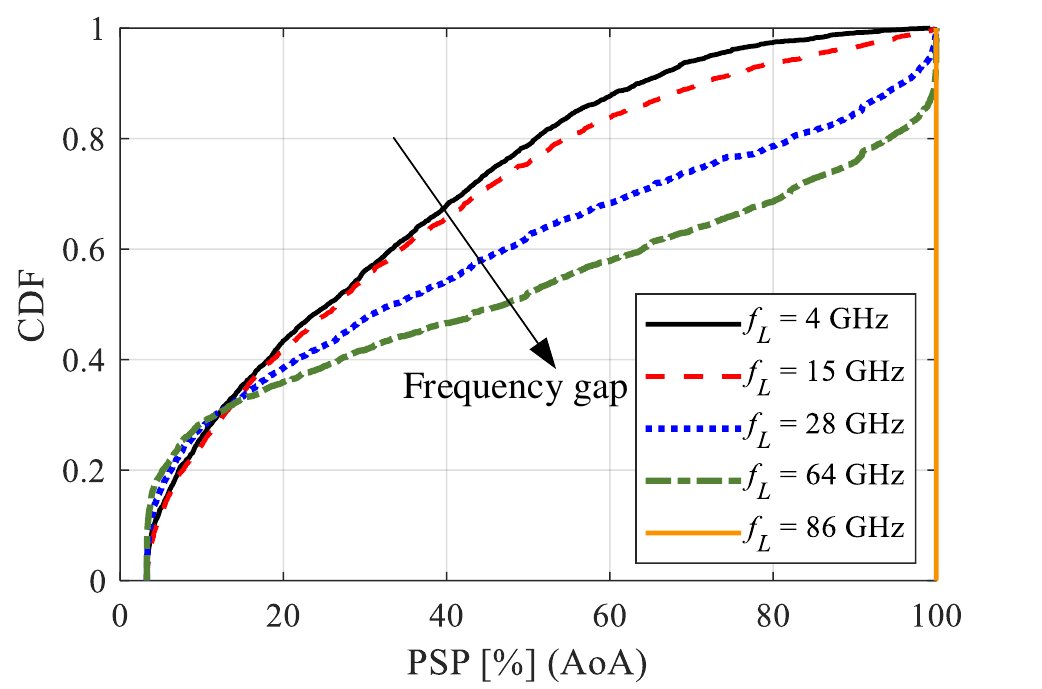}
	}
	\caption{(a) Normalized power patterns for PSP calculation (link 15). (b) Empirical CDF of the PSP metric with $f_U=$ 86~GHz and $f_L=$ 4, 15, 28, 64, 86~GHz.}
	\label{Fig:Metric1}
	\vspace{0pt}
\end{figure}

\subsubsection{Power Normalization}
Both estimated spectra are normalized to the sum power of unity such that they can be interpreted as probability distribution functions. The normalized PAS at lower frequency is calculated as
\begin{equation}\label{eq:NorPAS}
	B'_L(\alpha)=\dfrac{B_L(\alpha)}{\int B_L(\alpha)\mathrm{d}\alpha}
\end{equation}
The $B'_U(\alpha)$ is similarly defined with respect to the upper frequency PAS $B_U(\alpha)$. Fig. \ref{fig:PSP_2} depicts an example of normalized PAS $B'_L(\alpha)$ and $B'_U(\alpha)$ for link 15 at 4 GHz and 86 GHz, respectively. Compared with the results shown in Fig. \ref{fig:linkPAPDtop}, the dominant directions of wave propagation in azimuth plane are consistent at two frequencies, except for the path along the direction of 300$^{\circ}$.

\subsubsection{Total Variation Calculation}
As defined in \cite{Kyosti-18b}, the total variation distance between the normalized PAS is calculated as
\begin{equation}\label{eq:TVD}
	d_\mathrm{tv} = \dfrac{1}{2} \int \left| B'_L(\alpha) - B'_U(\alpha) \right| \mathrm{d}\alpha .
\end{equation}
The $d_\mathrm{tv}\in[0,1]$ characterizes the cumulative difference of normalized PAS between two frequency bands. Conversely, the PSP is defined as
\begin{equation}\label{eq:PSP}
	S = \left(1-d_\mathrm{tv} \right) \cdot 100\%
\end{equation}
The underlying example is the one illustrated in Fig. \ref{fig:linkPAPDside}, where the resulting PSP in this case is 56.99\%.

Empirical cumulative distribution function (CDF) of the PSP metric is collected over all 2875 links. In order to compare the spatial channel similarity across different frequency combinations, the CDF curves depicted in Fig. \ref{fig:PSP_all_cdf} are determined for five cases while keeping $f_U=$ 86 GHz and choosing $f_L=$ 4, 15, 28, 64, and 86~GHz. It can be observed that the similarity percentage increases in general with the frequency spacing between $f_U$ and $f_L$ decreasing. Ultimately, the similarity is always 100\% at $f_U=f_L$ as expected. On the other hand, we can remark that this trend does not hold for all curves and even reverses when the PSP values are below 13\%. An opposite trend but very close values shown in the low PSP region ($<$13\%) indicate that lower frequency channel can provide more rough angular information for higher frequency channel. However, low PSP value corresponds to significant difference between $B'_L(\alpha)$ and $B'_U(\alpha)$, which does not mean no accurate beam directions can be obtained from out-of-band channel. Moreover, the PSP metric only characterize the overall channel gain difference over all directions in 360$^\circ$ azimuth plane, lacking the details at specific directions. Hence, the PSP metric normally can be used to evaluate the channel similarity when multi-band channels showing good spatial congruence.

\subsection{Beam Direction-Based Similarity Metric}
\label{ssec:method2}
To avoid the reverse results at different levels of channel similarity and draw more practical conclusions, we define the other method with a focus on the collection of desired beam directions across different frequencies. Meanwhile, we also considers the assumed antenna size constraints at different frequencies, rather than using the same beampattern shown in Fig. \ref{fig:3gppBeam}.
We assume that low- and high-frequency radios and antenna systems operate equally at both link ends, where the lower frequency antenna system presumably has smaller electrical size, resulting in lower angular resolution and larger beamwidth of the main lobe. Next, two sets of best beam directions will be respectively identified at low and high frequencies, using the corresponding antenna beampatterns and propagation channel data. The sets of potential beam pointing angles in azimuth can be written as $\mathcal{A}_L=\{\theta_L|\theta_L \in [-\pi,\pi]\}$ and $\mathcal{A}_U=\{\theta_U|\theta_U \in [-\pi,\pi]\}$, respectively. Finally, we define a power ratio $R=\hat{P}_U(\theta_L)/\hat{P}_U(\theta_U)$ to measure how much potential channel gain is lost if the low-frequency beam direction information is used for high-frequency beam search instead of choosing its own channel information. Besides, we can count how many completely useless directions are provided from the low frequency for the high frequency.

\subsubsection{Determining Best Beam Directions}
\begin{figure}[!t]
	\centering
	\subfigure[][]{
		\label{fig:Balpha_2}
		\includegraphics[width=3.2in]{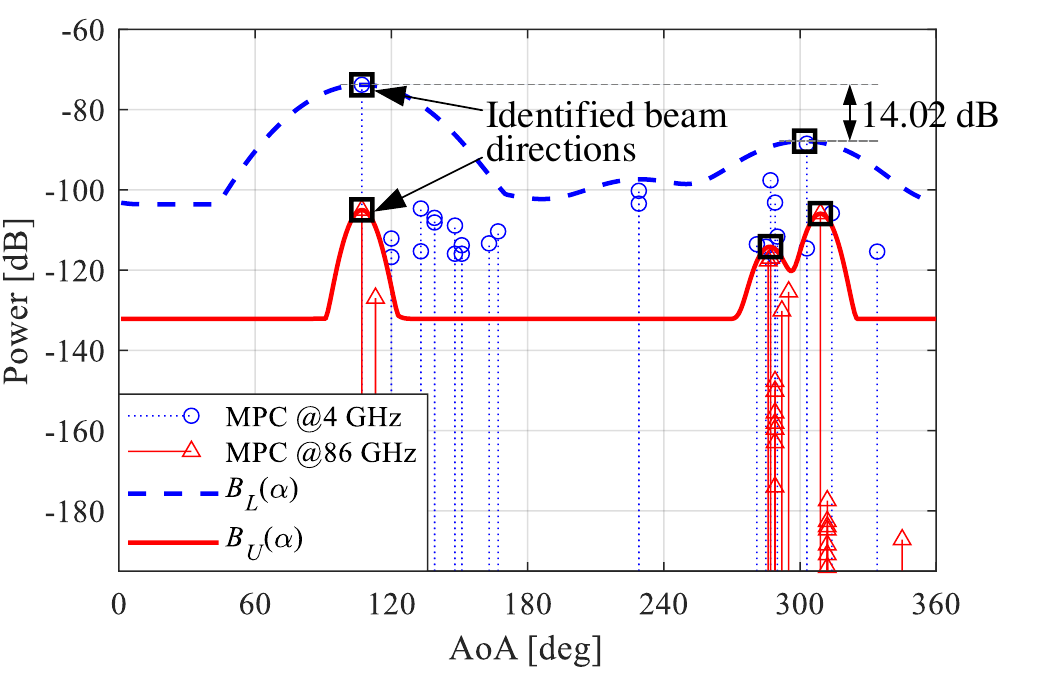}
	}
	\subfigure[][]{
		\label{fig:Rall_M1}
		\includegraphics[width=3.4in]{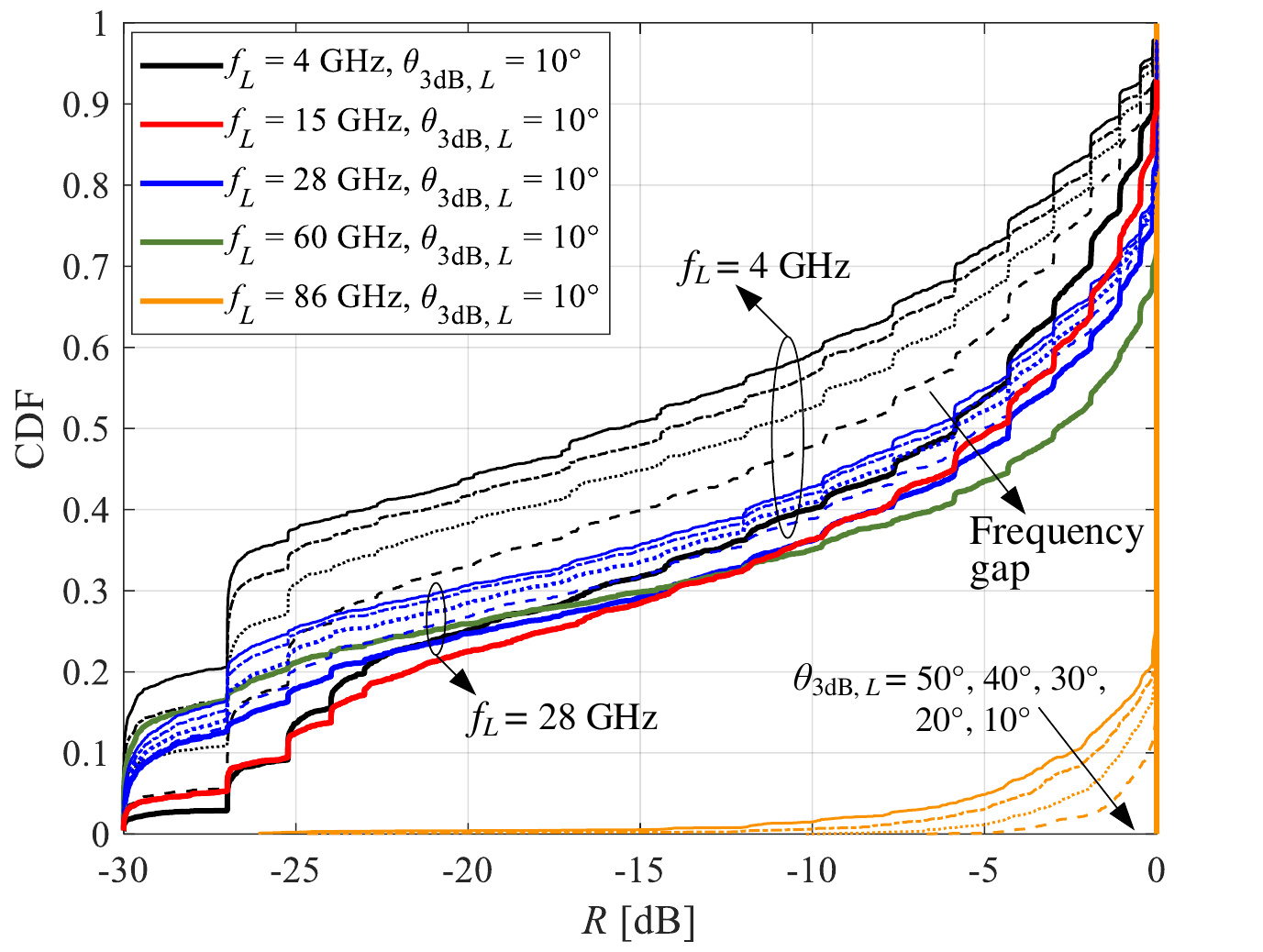}
	}
	\caption{(a) Original discrete PAS and filtered power patterns (link 15). (b) Empirical CDF of the power ratios $R$ under different $f_L$ and $\theta_{\mathrm{3dB},L}$ combinations with respect to $f_U=$  86 GHz and $\theta_{\mathrm{3dB},U}=10^{\circ}$.}
	\label{Fig:Metric2}
	\vspace{0pt}
\end{figure}

The sets of best beam directions $\mathcal{A}_L$ and $\mathcal{A}_U$ depend on the underlying radio channel, i.e. propagation channel and antenna systems. By the best beam directions we mean a minimum set of directions for beam steering or antenna orientation, such that the potential multipath channel gains can be maximally utilized. As is well known, the selection of best beam directions depends on tranceivers architectures, MIMO transmission schemes, beamforming strategies, etc., and cannot be exactly determined without the apriori knowledge. Thus, different gain patterns $G_L(\Omega)$ and $G_U(\Omega)$ are employed to filter the raw PADPs according to Eq. (\ref{eq:conv}). Fig. \ref{fig:Balpha_2} shows example filtered PAS $B_L(\alpha)$ and $B_U(\alpha)$ for link 15 using the 3GPP beam shape with the HPBW of 40$^{\circ}$ and 10$^{\circ}$ at 4 GHz and 86 GHz, respectively. In \cite{Kyosti-22a}, three methods are presented for identifying the desired directions. Here, the simplest method is utilized to choose the local maxima of $B(\alpha)$ with power threshold $\Delta_\mathrm{th}$ of 15 dB. As shown in Fig. \ref{fig:Balpha_2}, a total of two best beam directions (the black squares) are selected in lower band, while there are three directions in higher band. One more direction can be deteted in the direction of 295$^{\circ}$ at 86 GHz because of narrower beam enabling to distinguish observable directions without angle ambiguities.
\begin{figure*}[!t]
	\centering
	\subfigure[][]{
		\label{fig:ULApat}
		\includegraphics[width=2.25in]{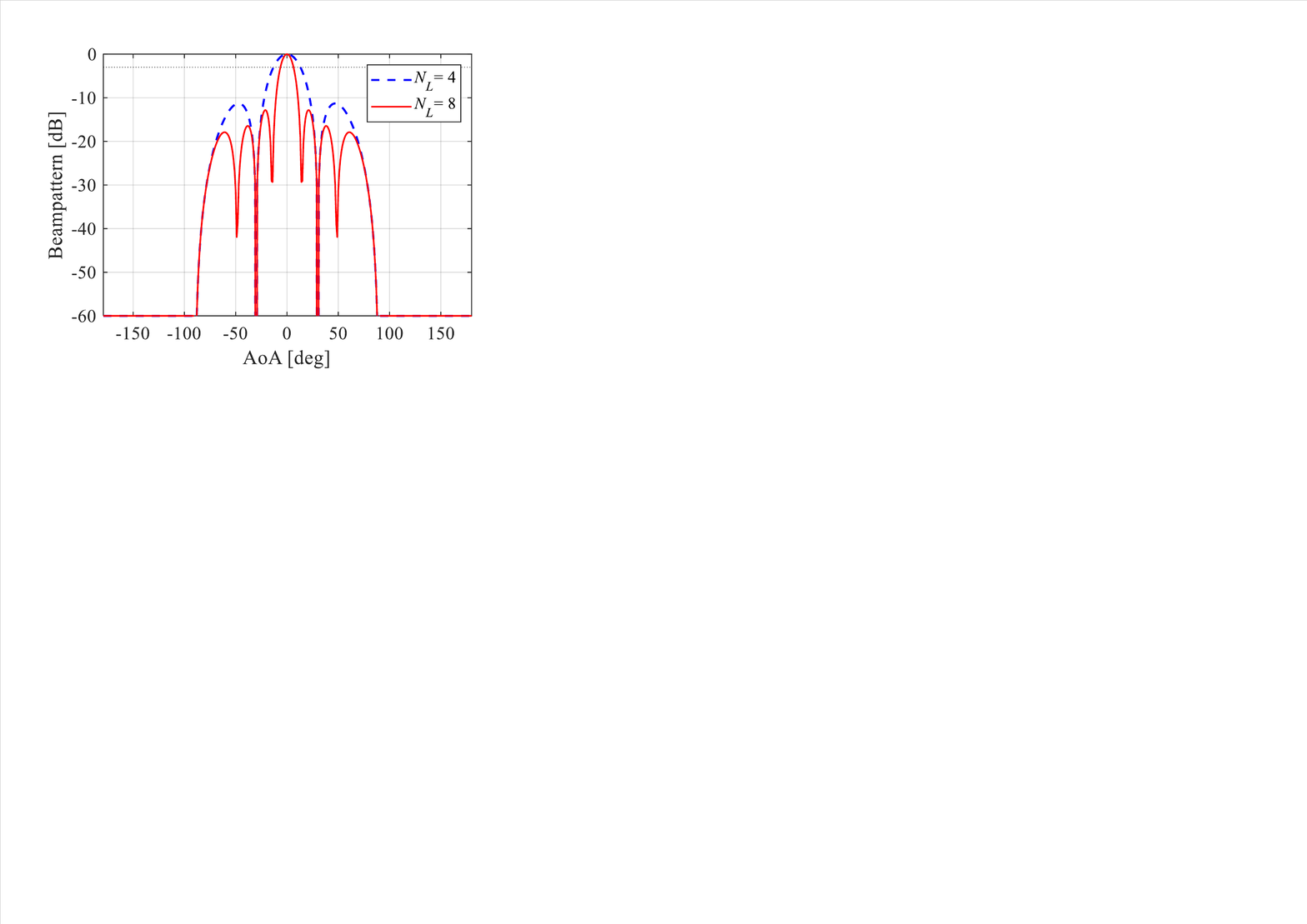}
	}
	\subfigure[][]{
		\label{fig:MeasCorridor}
		\includegraphics[width=2.25in]{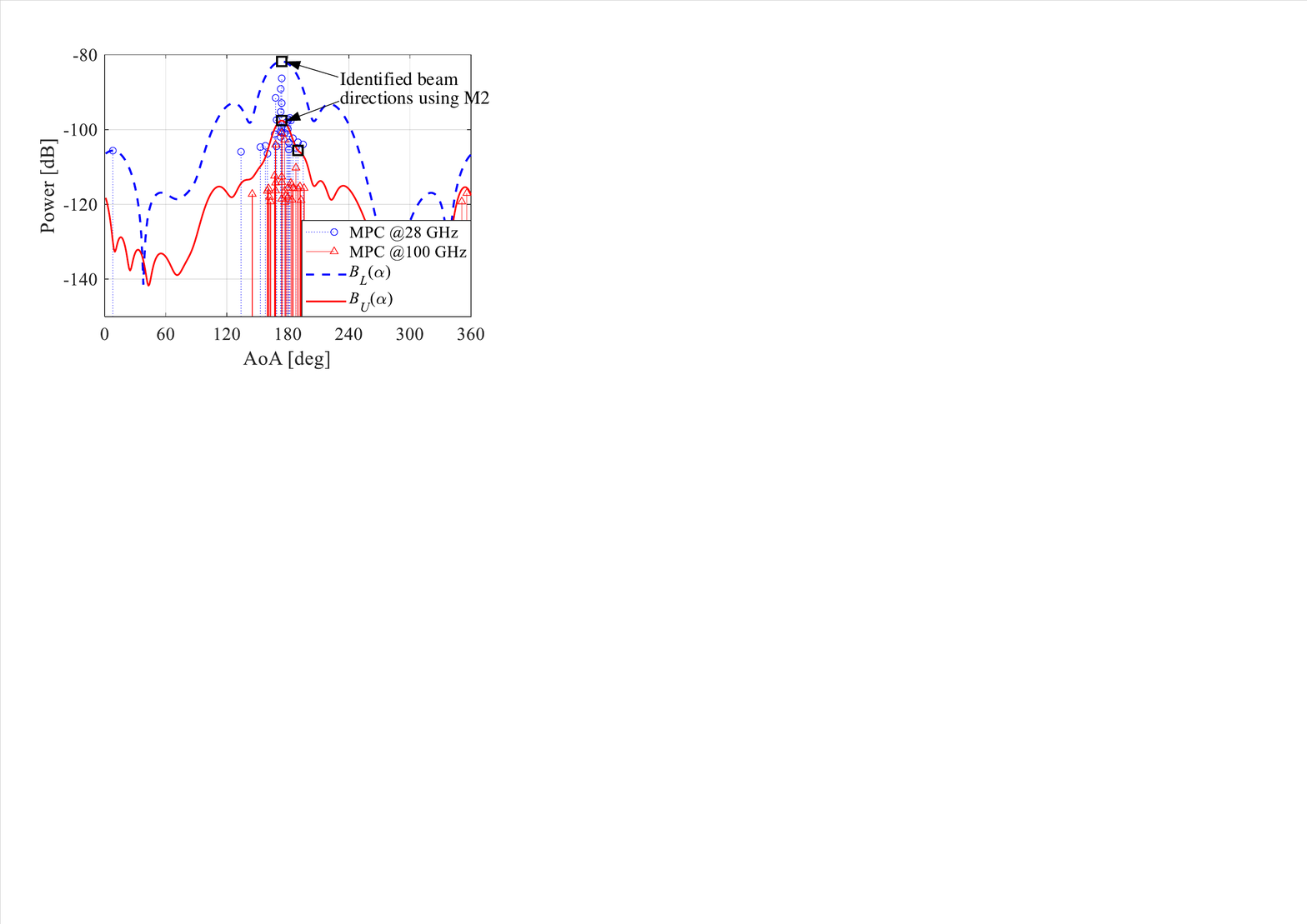}
	}
	\subfigure[][]{
		\label{fig:MeasEntrance}
		\includegraphics[width=2.25in]{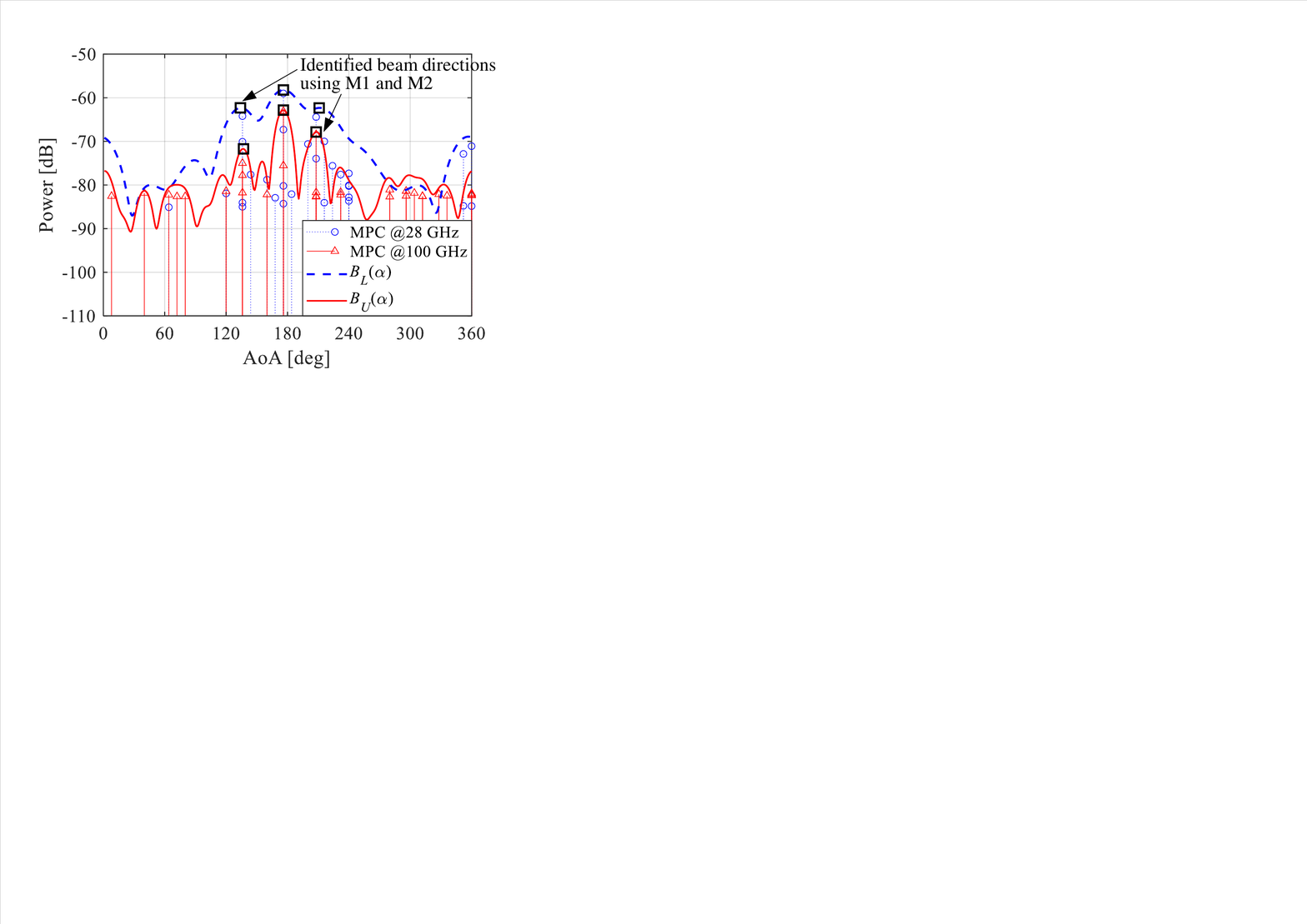}
	}
	\caption{(a) Beampatterns of 4- and 8-element ULAs. Measurement-based analysis of spatial channel similarity between 28 GHz and 100 GHz in (b) indoor corridor and (c) entrance hall.}
	\label{Fig:MeasComp}
	\vspace{-6pt}
\end{figure*}

\subsubsection{Calculating Power Ratio and False Direction Number}
Based on the estimated sets of best beam directions, we consequently define the power ratio as the summation of power observed by the high frequency antenna over different steering angle sets
\begin{equation}\label{eq:Pratio}
	R = \dfrac{\hat{P}_U(\theta_L)}{\hat{P}_U(\theta_U)} = \dfrac{\sum_{\theta_L \in \mathcal{A}_L} B_U(\theta_L)}{\sum_{\theta_U \in \mathcal{A}_U} B_U(\theta_U)}.
\end{equation}
The $R$ determines how much power is collected in the high-frequency case using different beam direction sets. Much larger $R$ indicates less power loss using low-frequency channel information and more obvious spatial channel similarity between two frequency.

A direction $\alpha \in \mathcal{A}_L$ proposed from low-frequency channel is interpreted as useless or "false" for high-frequency beam search if the high-frequency antenna does not collect any significant power when steered to that direction. More rigorously, we define the number of false directions as
\begin{equation}\label{eq:falseAlarm}
	N_f = \mathrm{card}\left( \mathop{\arg}\limits_{\alpha \in \mathcal{A}_L}
	10\lg\dfrac{B_U(\alpha)}{\mathop{\max}_{\alpha' \in \mathcal{A}_U} B_U(\alpha')} < \Delta_P
	\right),
\end{equation}
where $\mathrm{card}(\cdot)$ denotes the cardinality of a set and $\Delta_P$ is a power threshold (e.g., -30~dB). The number of these false directions can be counted per link and their empirical probability distribution over all links can be extracted. Likewise, the cardinality of sets  $\mathcal{A}_L$ and $\mathcal{A}_U$ can be collected, respectively, and their corresponding probability distribution function (PDF) can be extracted. For the case shown in Fig. \ref{fig:Balpha_2}, the corresponding $R$ and $N_f$ are respectively -2.02 dB and 0, while if $\Delta_\mathrm{th}$ is selected as 10 dB, $R$ and $N_f$ are -2.86 dB and 0, respectively.

It is evident that the power ratio $R$ depends on both the similarity of discrete PAS $P_L(\Omega)$ and $P_U(\Omega)$, as well as the shapes of antenna patterns $G_L(\Omega)$ and $G_U(\Omega)$. For assessing the proposed method, we determine the CDF of $R$ while keeping $f_U=$ 86~GHz and the HPBW of $G_U$ as 10$^{\circ}$, and taking selected combinations of $f_L=$ 4, 15, 28, 64, 86 GHz and the HPBW of $G_L$ as 50$^{\circ}$, 40$^{\circ}$, 30$^{\circ}$, 20$^{\circ}$, and 10$^{\circ}$. It can be observed from Fig. \ref{fig:Rall_M1} that the $R$ almost monotonically increases, i.e., the power loss at high frequency by using the low-frequency spatial channel information becomes smaller as the frequency gap decreases and the HPBW become much closer. The highest loss is 30~dB, which is determined by the peak to minimum gain ratio of the selected 3GPP beam shapes shown in Fig. \ref{fig:3gppBeam}. As expected, when the same frequencies (i.e., $f_L=f_U$) and HPBW are selected, the power loss is 0~dB over all links (see the orange bold solid line). Compared the CDFs of the $R$ versus the HPBW at 4 GHz and 28 GHz, there is a less obvious impact of beamwidth on channel similarity as the frequency gap decreasing. Moreover, it seems that the beamwidth has even stronger impact on channel similarity than the frequency difference. For example, the power ratio is much larger when using the spatial channel information at $f_L=$ 4 GHz with $\theta_{\mathrm{3dB}, L}=10^{\circ}$ in comparison with the results at $f_L=$ 28 GHz with $\theta_{\mathrm{3dB}, L}=$ 40$^{\circ}$.

\section{Results and Discussion}
\label{Sec:RandD}
Using the metric defined in Section \ref{ssec:method2}, the evaluation is performed on the feasibility of out-of-band spatial channel information for mmWave beam search based on the multi-band propagation data. Without loss of generality, frequency combinations (4, 86) GHz and (28, 100) GHz are chosen for simulation- and measurement-based analysis, respectively. We utilize more practical beampatterns of uniform linear arrays (ULAs) with different element number $N$ or (and) inter-element spacing at different frequencies, instead of ideal 3GPP beampattern.

\subsection{Validation with Dual-Band Measurements}
We choose a simple 4- and 8-element ULAs with 0.5 wavelength spacing for determining the $G_L(\Omega)$ and $G_U(\Omega)$, respectively. Beampatterns are defined for the $\Omega \in [-\pi/2,\pi/2]$ bore-sight half plane with all-zero phasing, while the back half plane $\Omega \in [-\pi,-\pi/2] \vee [\pi/2,\pi]$ is set to constant $-60$~dB gain. As shown in Fig. \ref{fig:ULApat}, the HPBW of main lobes are $26.2^{\circ}$ and $12.8^{\circ}$ for $G_L(\Omega)$ and $G_U(\Omega)$, respectively. The filtering of PAS by beampatterns is performed by rotating the patterns, not by setting various steering angles and using DFT weights, as would be the normal procedure with phased arrays. This approach is chosen since propagation paths are distributed over a wide sector of angles and in turn, we can remove the impact of ULA orientation on realized beam patterns. Two methods proposed in \cite{Kyosti-22a} are used. For method 1 (M1) each local maxima of $B(\alpha)$ within $\Delta_\mathrm{th}=10$~dB range from the global maximum is selected, and for method 2 (M2), the same 10~dB range is used, but instead of local maxima the directions are selected based on power and computational correlation between corresponding channel frequency responses.

As depicted in Fig. \ref{fig:MeasCorridor}, 1 and 2 beam directions can be extracted at 28 GHz and 100 GHz using M2, corresponding to 0.91 dB power loss, while only 1 and 1 beam directions can be extracted using M1 with power loss of 1.55 dB. In entrance hall (see Fig. \ref{fig:MeasEntrance}), 3 beam directions can be detected at two frequency bands with power loss of 0.13 dB and 0.09 dB using M1 and M2, respectively. The number of false directions are both 0 for these two cases. In general, M2 can estimate more possible beam directions compared with M1 if using narrower beampattern to filter propagation channel, which leads to smaller power loss. Measurement-based results show that our proposed metric performs well in LOS sceanrio between two bands with tens of gigahertz frequeny gap, especially when more advanced beam estimation method (e.g., M2) is adopted.

\subsection{Feasibility Study with Statistical Analysis}
The aforementioned results show significant impact of antenna HPBW on the measure of spatial channel similiarity across different frequencies using ideal 3GPP beampatterns. Here, we promote statistical analysis of radio channel similarity based on 4 GHz and 86 GHz ray-tracing simulation data, where more practical beampatterns are used to filter propagation channels.

\begin{figure}[!t]
	\centering
	\includegraphics[width=3.3in]{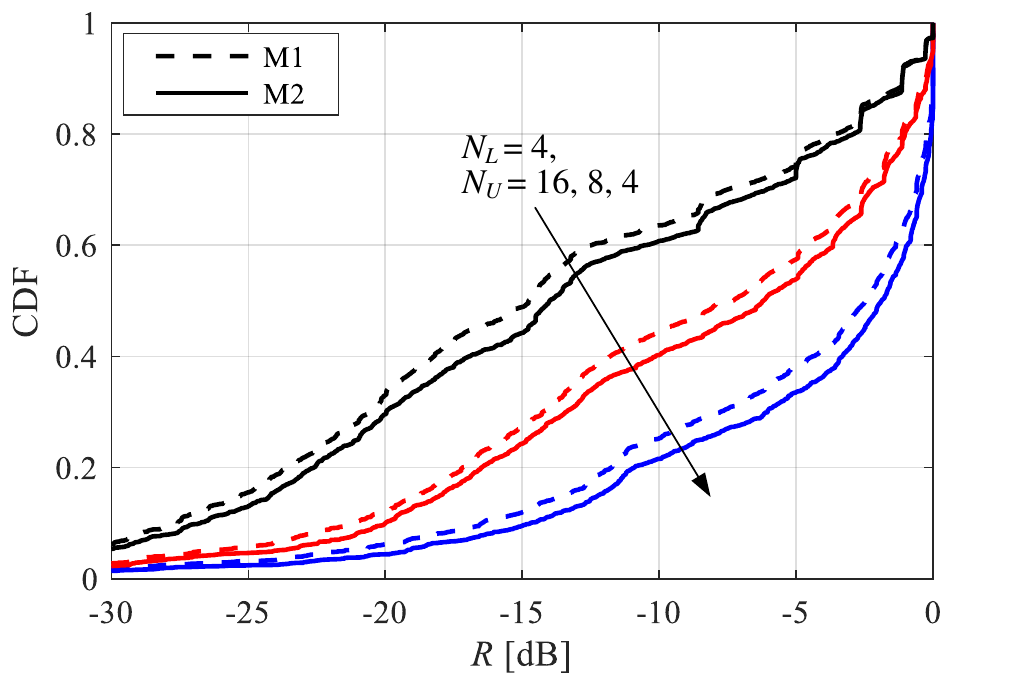}
	\caption{Empirical CDF of the power ratio metric based on ray-tracing simulation data.}\label{fig:Ractual}
	\vspace{0pt}
\end{figure}

The CDFs of power ratios $R$ over all 2875 links assuming different beampatterns and power ranges are shown in Fig. \ref{fig:Ractual}, as well as their statistics in Table \ref{tab:BeamStat}. It can be observed that the CDF curves for M2 can provide slightly higher values of $R$ compared M1. The values of $R$ are sensitive to the beamwidth. For example, using 4 GHz channel direction information to aid 86 GHz beam search, 2.27 dB and 7.31 dB power losses are introduced in 50\% of cases when 4- and 8-element ULAs are employed, respectively. Slightly increasing in $R$ can be observed if much larger $\Delta_\mathrm{th}$ is used during beam direction estimation. This is due to the fact that more channel directions can be detected with larger $\Delta_\mathrm{th}$, which potentially reduces the power loss between 4 GHz and 86 GHz radio channels. In the most common case with beampattern in Fig. \ref{fig:ULApat} and $\Delta_\mathrm{th}$= 10 dB, in 50\% of cases one would lose power 7.5 dB or less, if 86 GHz beam steering solely follows the out-of-band direction set $\mathcal{A}_L$ obtained from 4 GHz radio channel instead of the in-band direction set $\mathcal{A}_U$.

\begin{figure*}[!t]
	\centering
	\subfigure[][]{
		\label{fig:falsePDP4410}
		\includegraphics[width=2.25in]{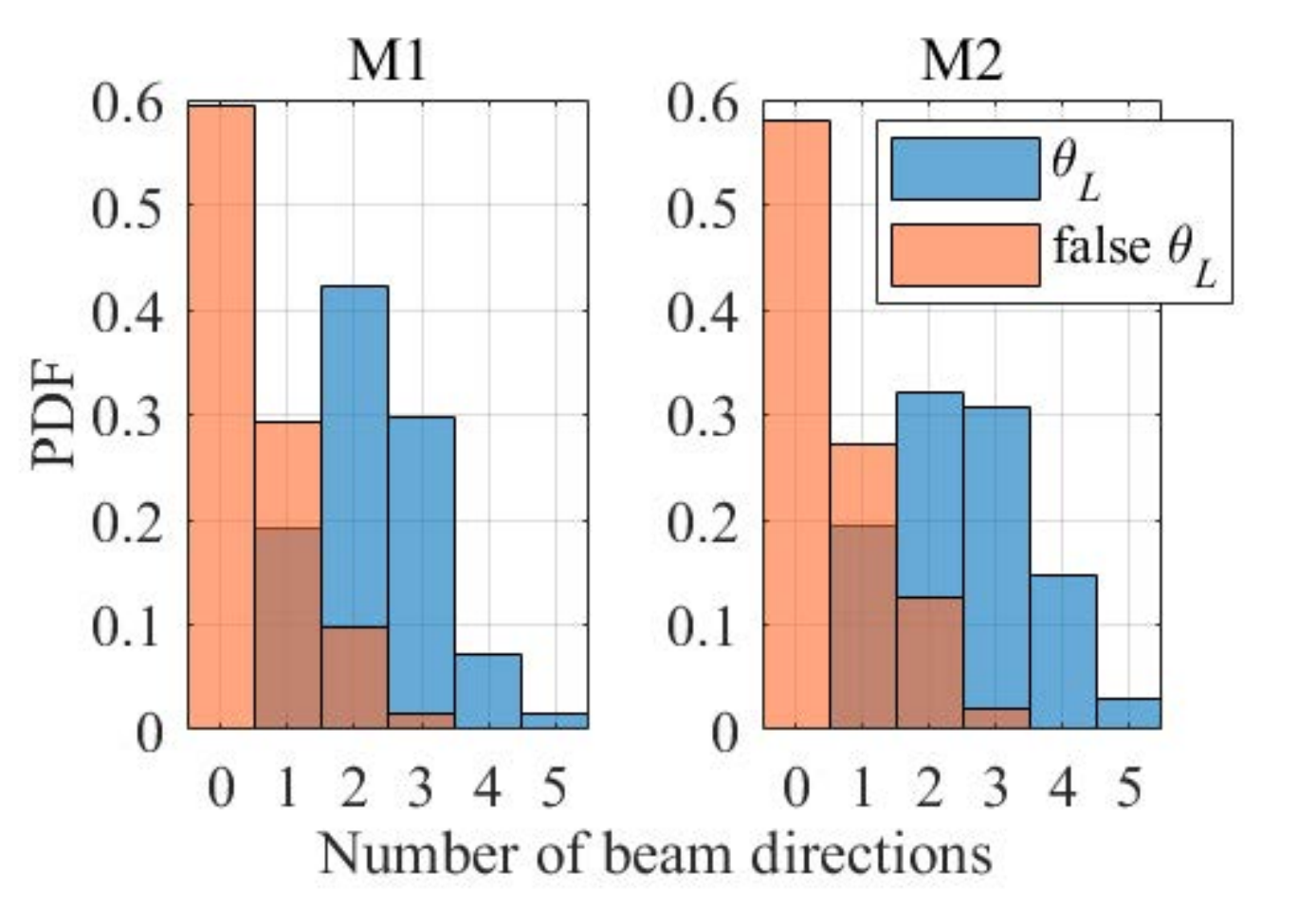}
	}
	\subfigure[][]{
		\label{fig:falsePDP4810}
		\includegraphics[width=2.25in]{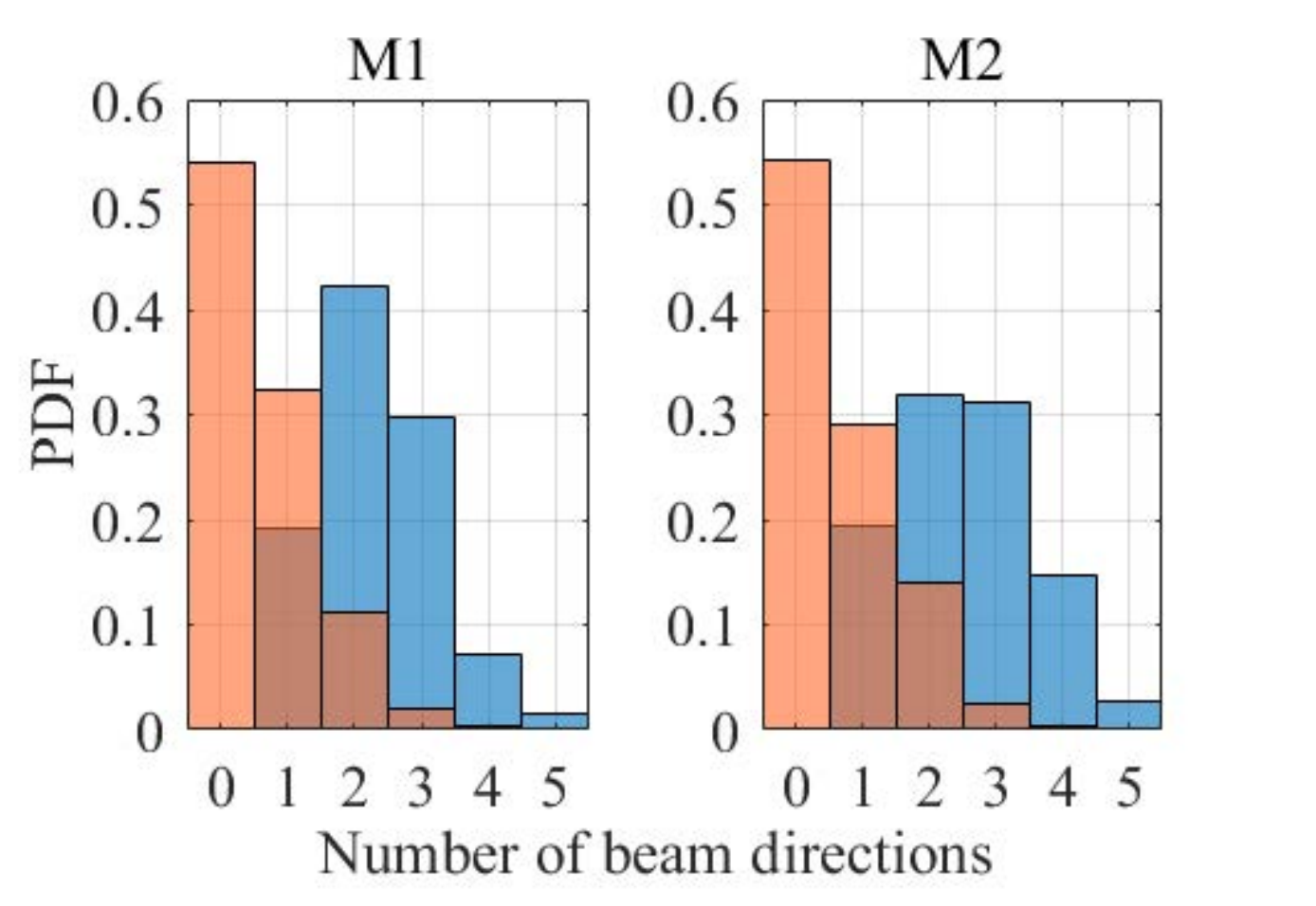}
	}
	\subfigure[][]{
		\label{fig:falsePDP4815}
		\includegraphics[width=2.25in]{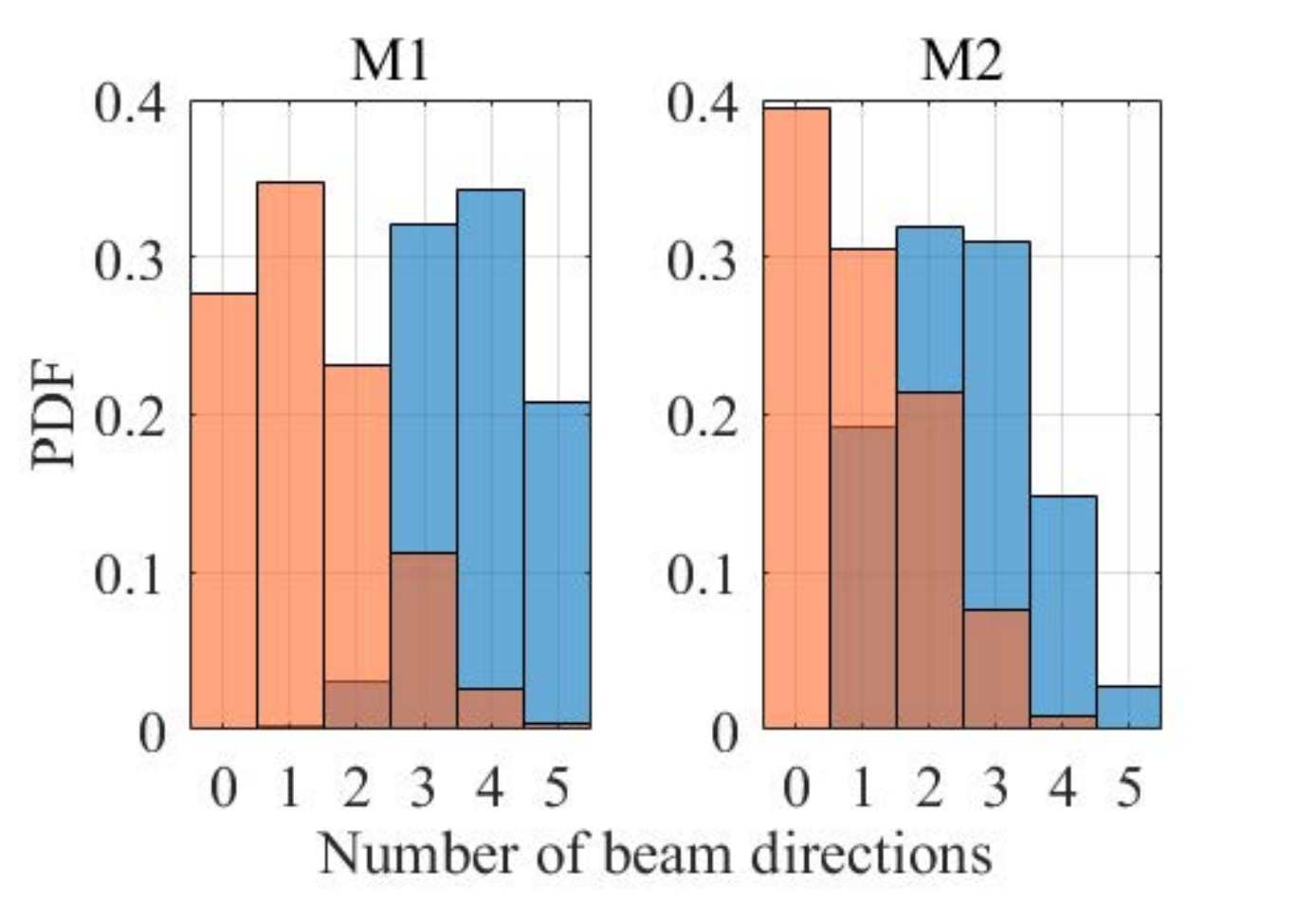}
	}
	\caption{Empirical PDF of the number of false directions and all extracted beam directions using 4 GHz out-of-band spatial channel information with the $(N_L, N_U)/\Delta_{\rm th}$ of (a) (4, 4)/10 dB, (b) (4, 8)/10 dB, and (c) (4, 8)/15 dB.}
	\label{Fig:falsePDP}
	\vspace{0pt}
\end{figure*}

While $R$ indicates whether all significant directions are included in 86 GHz beam steering, it is also important to consider how many false directions are included. Fig. \ref{Fig:falsePDP} depicts the PDFs of both the number of false directions $N_f$ and all identified beam directions within $\mathcal{A}_L$. Here, $\Delta_P$ in (\ref{eq:falseAlarm}) is set as -30 in this analysis. As expected, less false directions are introduced if the same beampatterns are used at 4 GHz and 86 GHz. For the case $(N_L, N_U)$= (4,8), despite less power loss observed for $\Delta_\mathrm{th}$= 15 dB, the probability of $N_f>$1 becomes much larger compared with the result for $\Delta_\mathrm{th}$= 10 dB. Hence, optimal selection of $\Delta_{\rm th}$ is necessary considering not all identified beam directions from lower band can provide sufficient channel gains in higher-frequency channels. We can conclude that the looser the criteria to find beam direction set $\mathcal{A}_L$, the better the power ratio $R$ and the worse the number of false directions $N_f$. There is a tradeoff between $R$ and $N_f$ when measuring the drgree of spatial similarity between low- and high-frequency radio channels. It is remarkable that $N_f$ is less than or equal to 1 in most of the cases ($>$80\%), meaning that spatial channel similarity between 4 GHz and 86 GHz radio channel can be exploited to realize coarse estimation of mmWave beam directions without noticeable accuracy decrease.
\begin{table}[!t]
	\centering
	\caption{The statistics of power ratios $R$ and number of false directions $N_f$ under different configurations.}
	\label{tab:BeamStat}
	\begin{tabular}{c|c|cc|cc|cc}
		\hline\hline
		\multicolumn{2}{c|}{$(N_L, N_U)/\Delta_{\rm th}$} & \multicolumn{2}{c|}{(4, 4) /10 dB} & \multicolumn{2}{c|}{(4, 8) /10 dB} & \multicolumn{2}{c}{(4, 8) /15 dB} \\
		\hline
		\multicolumn{2}{c|}{Method} & M1 & M2 & M1 & M2 & M1 & M2 \\
		\hline
		\multirow{3}{*}{$-R$ [dB]} & 10\% & 16.49 & 14.7 & 21.14 & 20.00 & 17.56 & 18.97 \\
		& 50\% & 2.27 & 1.88 & 7.31 & 6.20 & 6.26 & 5.08 \\
		& 90\% & 0 & 0 & 0.26 & 0.21 & 0.66 & 0.20\\
		\hline
		\multirow{2}{*}{\% $N_f$} & = 0 & 59.44 & 58.05 & 54.09 & 54.16 & 27.72 & 39.44 \\
		& $\leq$1 & 89.62 & 85.18 & 86.47 & 83.17 & 62.47 & 70.01 \\
		\hline\hline
	\end{tabular}
\vspace{0pt}
\end{table}

\subsection{Analysis of Unavailable Cases}
\begin{figure}[!t]
	\centering
	\includegraphics[width=3.3in]{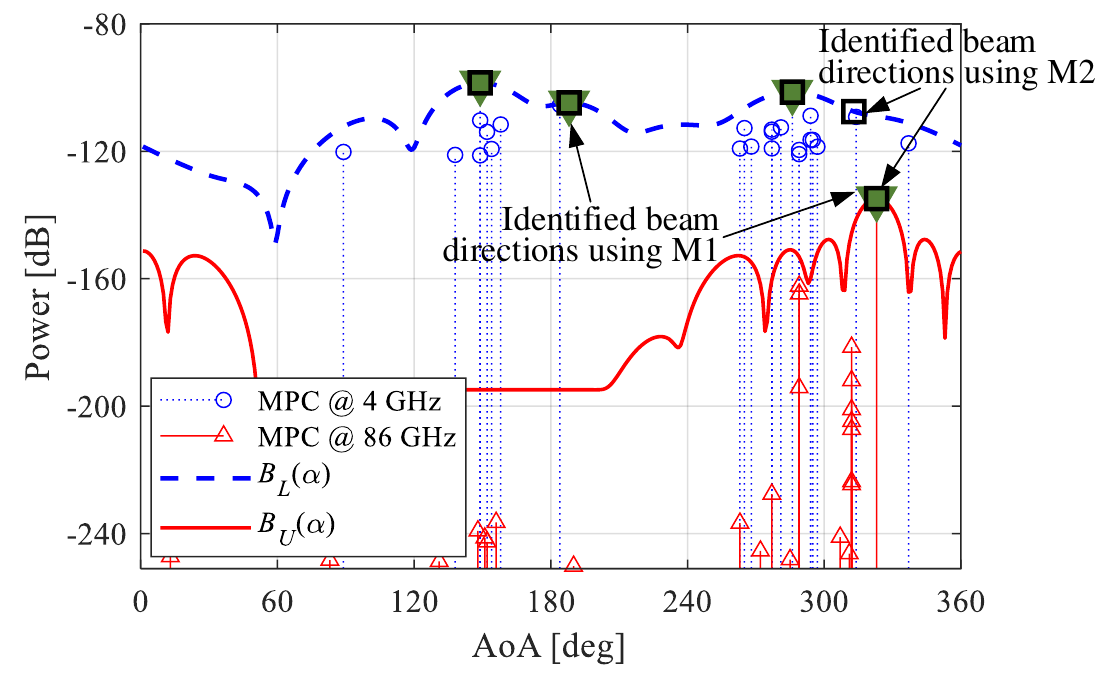}
	\caption{Original discrete PAS and filtered power patterns (Link 19).}\label{fig:Balpha19}
	\vspace{-6pt}
\end{figure}

There is a good share of cases where the 4~GHz beam information provides a good basis for the 86~GHz beam directions. A good example for link 15 is shown in Fig. \ref{fig:Balpha_2} with $N_f=0$ and $R=-2.86$ dB. However, there is another set of links where low-frequency radio channel filtered by smaller-aperture antenna provides no useful information for 86 GHz beam search. In general, all T-R links existing false beam directions, i.e., $N_f>0$, occurs along with smaller $R$. An example of such link is shown in Fig. \ref{fig:Balpha19} with $(N_L, N_U)$= (4,8), where $N_f=2$ (with M1) and $R=-16.15$ dB. Obviously, strong paths of 4 GHz channels do not coincide with those of 86 GHz channel, or more precisely, there is only a single dominant propagation path at 84 GHz that does not present at 4 GHz. Compared with ray-tracing and measured links exisitng strong and dominant paths, out-of-band spatial channel information cannot be leveraged for mmWave beam search in multipath rich environment, which easily leads to angle mismatch between two well-separated frequency bands.

\section{Conclusion}
\label{Sec:Conclusion}
In this paper, we conduct a feasibility study of out-of-band spatial information aided mmWave beam search from the radio channel similarity perspective. The beam direction-based spatial similarity metric is proposed, which compares the potential angular directions estimated from radio channels, as well as the power loss and number of false direction introduced by solely using out-of-band inforamtion. The quantitative analysis is performed based on point cloud ray-tracing and measured data. The results indicate that the usability of low-frequency radio channel information for high-frequency beam search evidently depends on many choices, e.g., of assumed angular resolutions and beampatterns of antennas, of the frequency gaps, of the criterion to select beam directions, and of accepting false directions. Overall, it seems practical to leverage available spatial channel information from lower frequency bands to aid mmWave beam steering from the radio channel point of view, but it should not rely on out-of-band information blindly.

For future work, the need of high-efficient beam search strategies becomes even more urgent in upper mmWave and terahertz bands (e.g., 100--300 GHz), since narrower beams increase the complexity of beam management under possible hardware limitations with potentially less flexibility. It would be of interest to include more statistical performance analysis of our proposed channel similarity metric with exhaustive directional propagation measurement campaigns. Moreover, developing a sophisticated metric for dynamic channel similarity measure is also an interesting research direction.


\bibliographystyle{IEEEtran}
\bibliography{Bibs/Bib_hexa,Bibs/Bib_pasi,Bibs/Bib_wei,Bibs/Bib_peize}

\end{document}